# Automated classification of *Hipparcos* unsolved variables

L. Rimoldini,[1,2]* P. Dubath,[1,2] M. Süveges,[1,2] M. López,[3] L. M. Sarro,[4] J. Blomme,[5]
J. De Ridder,[5] J. Cuypers,[6] L. Guy,[1,2] N. Mowlavi,[1,2] I. Lecoeur-Taïbi,[1,2] M. Beck,[1,2]
A. Jan,[1,2] K. Nienartowicz,[1,2] D. Ordóñez-Blanco,[1,2] T. Lebzelter[7] and L. Eyer[1]

[1]*Observatoire astronomique de l'Université de Genève, ch. des Maillettes 51, CH-1290 Versoix, Switzerland*
[2]*ISDC Data Centre for Astrophysics, Université de Genève, ch. d'Ecogia 16, CH-1290 Versoix, Switzerland*
[3]*Centro de Astrobiologia (INTA-CSIC), Departamento de Astrofisica, PO Box 78, E-28691, Villanueva de la Canada, Spain*
[4]*Dpt. de Inteligencia Artificial, UNED, Juan del Rosal, 16, E-28040 Madrid, Spain*
[5]*Instituut voor Sterrenkunde, KU Leuven, Celestijnenlaan 200D, B-3001 Leuven, Belgium*
[6]*Royal Observatory of Belgium, Ringlaan 3, B-1180 Brussel, Belgium*
[7]*University of Vienna, Department of Astronomy, Türkenschanzstrasse 17, A-1180 Vienna, Austria*

## ABSTRACT

We present an automated classification of stars exhibiting periodic, non-periodic and irregular light variations. The *Hipparcos* catalogue of unsolved variables is employed to complement the training set of periodic variables of Dubath et al. with irregular and non-periodic representatives, leading to 3881 sources in total which describe 24 variability types. The attributes employed to characterize light-curve features are selected according to their relevance for classification. Classifier models are produced with random forests and a multistage methodology based on Bayesian networks, achieving overall misclassification rates under 12 per cent. Both classifiers are applied to predict variability types for 6051 *Hipparcos* variables associated with uncertain or missing types in the literature.

## 1 INTRODUCTION

Stellar variability has opened a unique window into stellar physics and improved our understanding of phenomena such as stellar structure, evolutionary models, mass accretion, loss and transfer. Such knowledge has given rise to a wide range of applications, among which are asteroseismology (e.g. Aerts, Christensen-Dalsgaard & Kurtz 2010), the determination of the radius of stars in eclipsing binaries (even for objects as compact as white dwarfs; see Steinfadt et al. 2010), the measurement of distance with standard candles like Cepheids (e.g. Feast & Walker 1987) or RR Lyrae (see Smith 2004, for a comprehensive account) and the exploitation of RR Lyrae stars to probe the dark matter halo (Sesar et al. 2010) or constrain the neutrino magnetic moment (Castellani & Degl'Innocenti 1993). Non-periodic variable sources have also contributed to improving our understanding from cosmic scales, like the expansion of the Universe employing supernovae as distance indicators (e.g. Riess et al. 1998; Perlmutter et al. 1999) and dark matter constraints with microlensing (e.g. Alcock et al. 2000), down to phenomena on stellar scales, such as variations associated with circumstellar discs (e.g. Mennickent, Sterken & Vogt 1997), flares and other chromospheric activities (see Eker et al. 2008, for a catalogue). As the number of identified variables and measurement quality increase, statistical analyses become possible, thereby extending the reach of other methods.

The large data volumes in recent and future astronomical surveys, such as the Sloan Digital Sky Survey[1] (SDSS, Ahn et al. 2012), the Panoramic Survey Telescope & Rapid Response System[2] (Pan-STARRS, Kaiser et al. 2002), Gaia[3] (Perryman et al. 2001) and the Large Synoptic Survey Telescope[4] (LSST, Ivezić et al. 2011), necessitate automated procedures to analyse and classify variable stars. A citizen science project such as Galaxy Zoo[5] (Lintott et al. 2008) pioneered the morphological classification of galaxies by involving the online community, leading to over 50 million reliable classifications of galaxies in a year thanks to the participation of over $10^5$ volunteers. While such efforts are well suited to recognize patterns difficult to extract automatically, the development, improvement and tuning of automated procedures remain fundamental for efficient processing of the bulk of the data.

Supervised classification is a specific branch of data mining which aims at the identification of classes (or hereafter 'types') of objects inferred from a set of other objects (identified as the training set) of known types. In the case of variable stars, such types group sources according to their cause of variability, whether intrinsic or extrinsic.

---

[1] http://www.sdss.org
[2] http://pan-starrs.ifa.hawaii.edu/public
[3] http://gaia.esa.int
[4] http://www.lsst.org/lsst
[5] http://www.galaxyzoo.org

*E-mail: lorenzo@rimoldini.info

One approach to the automated supervised classification of variable stars involves the characterization of sources with 'attributes' describing the context, features, statistical properties and the time series modelling. Attributes do not probe directly the cause of variability, but provide a set of measures to assess the most likely variability type statistically. The classification procedure passes attributes to machine-learning algorithms for training the classifier, so that the trained model can predict types for unclassified objects. This methodology was applied to classify variable stars into several periodic types with data from various surveys by a number of recent studies (Debosscher et al. 2007; Willemsen & Eyer 2007; Sarro et al. 2009; Blomme et al. 2010, 2011; Dubath et al. 2011; Richards et al. 2011), which employed machine-learning techniques such as random forests (Breiman 2001), classification and regression trees (Breiman et al. 1984), boosted trees (Freund & Schapire 1996), Gaussian mixture models (McLachlan & Basford 1988), Bayesian networks (Pearl 1988), Bayesian average of artificial neural networks (Neal 1996) and support vector machines (Gunn, Brown & Bossley 1997).

Dubath et al. (2011) described an automated classification procedure applied to variable stars from the *Hipparcos* periodic catalogue (ESA 1997), which had sufficient information to derive a period or confirm the one from the literature. Hundreds of *Hipparcos* sources from the periodic catalogue with still uncertain or missing classification were classified employing a classifier trained on a subset of known periodic sources. The present work focuses on those objects whose variability has been inferred from the *Hipparcos* data, but periodicity could not be determined independently with the *Hipparcos* data alone nor confirmed using periods available in the literature at the time of publication of the *Hipparcos* catalogue (1997). The scope of Dubath et al. (2011) is broadened by including the classification of sources manifesting non-periodic and irregular light variations and extending to regimes of low signal-to-noise (S/N) ratios. Predictions of variability types are provided for thousands of sources from the *Hipparcos* unsolved catalogue and a set of microvariables, most as yet unclassified.

The detection of variable sources from survey data often generates a relevant fraction of apparently quasi-periodic and non-periodic objects, due to intrinsic properties or low S/N levels, in addition to a set of clean periodic variables. The organization of the *Hipparcos* variables into separate catalogues based on a combination of periodicity, S/N ratio and amplitude proved useful for follow-up studies such as this. Despite the context of data characterized by low S/N ratios, we found sufficient information for classification goals and identified a significant number of stars exhibiting irregular light variations, whether due to flares, eruptions, superposition of multiple close periods or other phenomena. Such cases included long-term light variations with poorly defined or occasional periodicity (typically giants or supergiants of late spectral types and slow red irregular variables), transient phenomena like eruptive variables interacting with forming discs (of the Gamma Cassiopeiae type), non-uniform surface brightness, flares and chromospheric activity (as in BY Draconis and RS Canum Venaticorum stars). Our classification results are listed in a catalogue available online, from which objects of interest can be selected.

This paper is organized as follows. The *Hipparcos* data subset employed herein is described in Section 3; the variability types and the training set of sources for supervised classification are presented in Section 4; the most relevant attributes are defined in Section 5; the random forest classification method, the attribute importance, the classification strategies and results are described in Section 6; a multistage approach on the same data with Bayesian networks is presented in Section 7; the set of sources for which variability types are predicted and the associated classification results are defined in Section 8; and our final conclusions are reported in Section 9.

## 2 TERMINOLOGY

Classification results are assessed, among others, through quantities derived from confusion matrices, which summarize the outcome of the classification of known test objects by listing the numbers of true versus predicted variability types in rows and columns, respectively (as in Fig. 4). Such indicators of classification quality are defined herein for clarity. Assuming that a test set contains $R$ sources of a given type (corresponding to a row of the confusion matrix) and $C$ objects are classified as such a type (summing the number of predictions in the corresponding column of the confusion matrix), we denote the numbers of true positives, false positives and false negatives by TP, FP and FN, respectively, and define the following:

   (i) the accuracy (or completeness, efficiency) rate as TP/$R$,
   (ii) the precision (or purity) rate as TP/$C$,
   (iii) the misclassification (or error) rate as FN/$R$ or $1-$(TP/$R$) and
   (iv) the contamination rate as FP/$C$ or $1-$(TP/$C$). If no sources are classified as a certain type (i.e. $C = 0$), the contamination rate is not determined for that type.

## 3 THE *HIPPARCOS* UNSOLVED VARIABLES

The *Hipparcos* mission (ESA 1997) provides astrometric and photometric data for the brightest sources in the sky. The *Hipparcos* catalogue (Perryman et al. 1997) contains 118 204 entries with associated photometry, among which 11 597 sources are identified as variable: 2712 and 5542 of them were published in the periodic and unsolved catalogues (ESA 1997, Vol. 11, Parts 1 and 2), respectively, and 3343 objects (comprising other unsolved sources and microvariables) remained uninvestigated.

The sources not included in the *Hipparcos* periodic catalogue were originally distinguished in two sets.

   (i) Unsolved variables (U): 7782 sources with amplitudes greater than 0.03 mag. They included periodic and non-periodic sources, transient phenomena, suspected variables and high-confidence microvariables.[6]

   (ii) Microvariables (M): 1045 sources with a $\chi^2$ probability of being constant of less than $10^{-4}$, amplitudes smaller than 0.03 mag and without established periods, which remained uninvestigated by ESA (1997). High-confidence microvariables were moved to the unsolved catalogue.

In order to include the information embedded in the *Hipparcos* organization of sources, possibly hinting at different levels of confidence in the classification results, this work separates the sources in set U as follows.[6]

---

[6] Sets U and $U_1$ exclude 55 sources flagged with 'R' (indicating calibration using an erroneous colour index) and 1 source flagged with 'D' (suggesting duplicity or photometric disturbances) in the field H52 of the *Hipparcos* catalogue (ESA 1997). Also, sources HIP 65908 and HIP 111858 were excluded (and removed from sets U and $U_2$) because they lacked precise positional indicators (fields H8 and H9) needed for cross-correlations with other catalogues.

**Table 1.** Variability types included in the training set are listed together with the corresponding number of instances and references. Labels in bold indicate types which exhibit light variations with intrinsic irregularities or apparent non-periodicity, either exclusively or occasionally, while the other types are periodic. The training set contains 3881 sources in total. The numbers of objects from the *Hipparcos* periodic catalogue, selected by Dubath et al. (2011)[a], are listed in column 'P', while those from the unsolved catalogue (U) and a set of microvariables (M) are included in column 'X' (X = U+M).

| Type | Label | P | X | Reference |
|---|---|---|---|---|
| Irregular | ***I*** | – | 54 | Watson et al. (2011) |
| Long period variables | ***LPV =*** | 285 | | Lebzelter (private communication) |
| (Mira Ceti, semiregular & slow irregular) | *M + SR + L* | | 1464 | Watson et al. (2011) |
| RS Canum Venaticorum | ***RS + BY*** | 35 | | Eker et al. (2008) |
| & BY Draconis | | | 138 | Watson et al. (2011) |
| B-type emission line star | ***BE + GCAS*** | 13 | 249 | Watson et al. (2011) |
| & Gamma Cassiopeiae | | | | |
| Slowly pulsating B-type star | *SPB* | 81 | – | De Cat (private communication) |
| Alpha$^2$ Canum Venaticorum | *ACV* | 77 | | Romanyuk (private communication) |
| | | | 27 | Watson et al. (2011) |
| Eclipsing binary: Algol type | *EA* | 228 | | ESA (1997) |
| | | | 131 | Watson et al. (2011) |
| Beta Lyrae type | *EB* | 255 | – | ESA (1997) |
| W Ursae Majoris type | *EW* | 107 | – | ESA (1997) |
| Ellipsoidal rotating | *ELL* | 27 | – | ESA (1997) |
| Alpha Cygni | ***ACYG*** | 18 | 35 | Watson et al. (2011) |
| Beta Cephei | *BCEP* | 30 | | De Cat (private communication) |
| | | | 26 | Watson et al. (2011) |
| Delta Cephei | *DCEP* | 189 | – | Watson et al. (2011) |
| First overtone | *DCEPS* | 31 | – | Watson et al. (2011) |
| Multimode | *CEP(B)* | 11 | – | Watson et al. (2011) |
| RR Lyrae: asymmetric light curve | *RRAB* | 72 | – | Watson et al. (2011) |
| Nearly symmetric light curve | *RRC* | 20 | – | Watson et al. (2011) |
| Gamma Doradus | *GDOR* | 27 | | De Cat (private communication) |
| | | | 17 | Watson et al. (2011) |
| Delta Scuti (incl. SX Phoenicis) | *DSCT* | 47 | 20 | Watson et al. (2011) |
| Low amplitude | *DSCTC* | 81 | 61 | Watson et al. (2011) |
| W Virginis: period > 8 d | *CWA* | 7 | – | Watson et al. (2011) |
| Period < 8 d | *CWB* | 6 | – | Watson et al. (2011) |
| SX Arietis | *SXARI* | 7 | – | Romanyuk (private communication) |
| RV Tauri | *RV* | 5 | – | Watson et al. (2011) |
| | Total: | 1659 | + 2222 = 3881 | |

[a]See footnote 10 in the text.

(i) 'U$_1$': 5486 unsolved variables included in the variability annex (ESA 1997, Vol. 11, Part 2).

(ii) 'U$_2$': 2296 unsolved variables not included in the variability annex, which remained uninvestigated at the time of publication of the *Hipparcos* catalogue.

The Variable Star Index[7] (Watson, Henden & Price 2011) of the American Association of Variable Star Observers (AAVSO) was employed for up-to-date literature information[8] on the variability types associated with the *Hipparcos* sources, which were cross-matched with the nearest object in the AAVSO catalogue, mostly within a few arcsec.[9]

# 4 TRAINING SET

The training set for supervised classification includes sources from the *Hipparcos* periodic, unsolved and microvariable sets. Table 1 lists the variability types considered in this work, their labels (adopted hereafter) and the corresponding number of representatives. The training set is composed of 3881 time series: 1659 periodic variables are selected from the *Hipparcos* periodic catalogue according to Dubath et al. (2011)[10] and 2222 sources originate from the *Hipparcos* unsolved catalogue and a set of microvariables. The group of 2222 training-set sources is specific to this work; it includes objects cross-matched within 1 arcsec from the *Hipparcos* sources and with secure classifications in the AAVSO catalogue (Watson et al. 2011), i.e. labelled without the uncertainty symbol ':'. The training set excludes sources with types which are unknown (like future new types or brief transitions between variability types), uncertain, generic (such as periodic, pulsating, double, pre-main-sequence, miscellaneous or variable), hastily designated (not confirmed by further observations), unstudied (with slow or rapid light variations) and possibly alternative or combined. This first selection was based on the variability type labels provided by Watson et al. (2011).

---

[7] http://www.aavso.org/vsx/index.php
[8] The Variable Star Index of the AAVSO includes the most recent information from the General Catalogue of Variable Stars (Samus et al. 2012) and selected variables from other surveys, such as the Northern Sky Variability Survey (Woźniak et al. 2004), the All Sky Automated Survey 3 (Pojmanski 2002), the Optical Gravitational Lensing Experiment 2 (Szymański 2005), and new variables reported in the Information Bulletin on Variable Stars. Additions of new variables and modifications of existing records are peer reviewed and referenced. More information on the compilation of the AAVSO database and on how it is maintained and revised is available at http://www.aavso.org/vsx/index.php?view=about.top

[9] About 95 per cent of sources in sets U and M were cross-matched with the AAVSO catalogue within 1 arcsec, with some variations depending on the set of provenance: over 95 per cent of sources from set U$_1$ were cross-matched within 1 arcsec, over 90 per cent of sources from set M had counterparts within 2 arcsec and over 80 per cent of sources from set U$_2$ were associated with AAVSO objects within 5 arcsec.

Additional requirements were applied to the selected secure types, and further sources from the *Hipparcos* unsolved catalogue were excluded in the following cases.

(i) Extremely weak representatives (like noisy eclipsing binaries with measurements during eclipses consistent with non-eclipsed data within $\approx 3\sigma$).

(ii) Very poorly sampled sources (significantly more sparse than average, e.g. with a gap covering over half of the cycle in the plot of the folded time series).

(iii) Time series with less than 30 reliable measurements.

(iv) Variability types with less than 12 secure representatives.[11] A minimum number of representatives were not required for types characterized by sources from the periodic catalogue, which benefited from greater S/N ratios and clear periodicity.

Considering the challenges posed by the characterization of irregular features and the lower S/N levels of the *Hipparcos* unsolved catalogue and a set of microvariables with respect to the periodic sample, objects with missing information (like colour or parallax not available in the catalogue) were excluded from the training set, and only good quality measurements were considered [with quality flag HT4 of ESA (1997) not set, or only bit 0 set]. We did not employ methods to deal with sources with missing information (e.g. see Stekhoven & Bühlmann 2012) because of the small number (19) of objects concerned, which typically depended mostly on the missing attribute and were already well represented in the training set.

When the available data were considered insufficient a priori to separate certain variability types, they were deemed fairly confused and grouped together. For example, Delta Scuti types are phenomenologically close to SX Phoenicis variables, but since they are stars of Populations I and II, respectively, they could be distinguished employing indicators of metallicity. Since such information was not available in our data, SX Phoenicis stars were incorporated in the *DSCT* type. Also, slow irregulars included sources with long-term variations, exhibiting (or not) occasional periodic features:

many of them were often insufficiently studied and were actually semiregulars, so they were merged with the long period variables. Other types which might combine periodic and irregular light variations involved transient phenomena occurring in *BE* +*GCAS* and *RS* +*BY* variables, while *ACYG* stars might appear irregular because of the superposition of multiple close periods (which lead to oscillation beats). Irregular variables (type *I*) formed a coarse group of objects with unknown light variations and irregularities, in need of further study, and they might actually belong to other types.

Variability types which appear to contain non-periodic variations (whether exclusively or occasionally) are listed in bold font in Table 1. The semi-regular (*SR*), slow irregular (*L*) and irregular (*I*) groups include several subtypes (e.g. *SRA*, *SRB*, *SRC*, *SRD*, *SRS*, *LB*, *LC*, *IA*, *IB*, *INA*, *INB*). The reader is referred to the General Catalogue of Variable Stars (GCVS, Samus et al. 2012)[12] for a detailed taxonomy of variable stars and description of several variability types and subtypes.

Suffixes indicating the *Hipparcos* catalogue of origin were added to the variability type labels: 'P' for sources selected from the periodic catalogue and 'X' for sources extracted from the unsolved catalogue and a set of microvariables (owing to the definitions introduced in Section 3, X = U+M). This distinction gave the opportunity to benefit from additional information based on periodicity detection and S/N ratio, but it should only be regarded as an added value to the classification results. The accuracy of classification was evaluated regardless of the confusion between label suffixes 'P' and 'X' for the same type.

Some of the variability types could only be represented by objects with high S/N ratios and reliable periods (from the periodic set). While this retains the accuracy of the class description, the different distribution (and in particular coverage) of training-set attributes, with respect to those of sources in the prediction set, might cause sample selection biases. Methods to improve the training set in such circumstances have recently been proposed by Richards et al. (2012) and Long et al. (2012), although these have not been implemented herein.

## 5 CLASSIFICATION ATTRIBUTES

Our initial set of attributes contained about 150 elements (originated from catalogues, modelling results, statistical analyses) which explored various ways to describe sources and characterize their time series and associated features. Eventually, only the most relevant attributes for classification were retained (see Section 6.2). Their selection followed the iterative procedure described by Dubath et al. (2011), which also limited the Spearman correlation coefficients between any pair of attributes to values under 80 per cent. Reducing the dimensionality of attribute space not only made the machine-learning process easier and quicker, but it also reduced the confusion created by many irrelevant attributes and possible complex interactions with the truly useful ones. This is expected to improve the robustness of the classifier and produce models which are easier to interpret.

Catalogue information included average colours, parallax and location with respect to the Galactic plane. Near-infrared photometry from the Two Micron All Sky Survey (Skrutskie et al. 2006) was highly correlated with the optical bands and was not selected in the list of the most important attributes.

---

[10] Only two representatives of the *CWA* type were removed from the training set of Dubath et al. (2011) because the recovered periods were smaller than 8 d, contrary to the definition of this type.

[11] Variability types which have been excluded, because represented by too few representatives, include (with numbers in parentheses): Wolf–Rayet variables (11), R Coronae Borealis (10), S Doradus (5), FK Comae Berenices (5), Novae (5), Z Andromeda (4), X-ray variables with optical counterparts (4), UV Ceti (3) and PV Telescopii (3). Unfortunately, not a single instance of these classes could be classified correctly when included in the training set, as expected from the combination of small numbers and high noise levels, and thus they have not been considered herein.

[12] http://www.sai.msu.su/gcvs/gcvs/iii/vartype.txt

We employed time series measured in the *Hipparcos* band (van Leeuwen et al. 1997) and modelled them with Fourier series as described by Dubath et al. (2011).

### 5.1 Attribute description

The most important attributes for classification are described below in order of importance (units are denoted in brackets, when applicable).

(1) '*V* − *I*' (mag): the reddened *V* − *I* colour index in Cousins' system, as provided by ESA (1997).

(2) '*Skewness*': the unbiased skewness of the distribution of *Hipparcos* magnitudes, weighted by the inverse of squared measurement uncertainties. Weighting improved the importance of this attribute because the skewness proved especially useful for identifying eclipsing binaries, and the weighted mean (around which the third moment was computed) fitted the bulk of non-eclipsed data better than the unweighted counterpart. Such weighting is effective when errors are expressed in terms of magnitudes, because generally smaller errors correspond to brighter (non-eclipsed) data.

(3) '*Amplitude*' (mag): the difference between the faintest and the brightest values of the light-curve model.

(4) '*Period*' (day): the most likely period computed with the generalized Lomb–Scargle method (Zechmeister & Kürster 2009) for sources with weighted skewness of the magnitude distribution smaller than 1.6. Larger values of skewness were typical of eclipsing binaries, and the use of weights in period search (as the inverse of squared measurement uncertainties) removed importance to the eclipsed data containing the periodic signal. Thus, periods of sources with *Skewness* > 1.6 were computed with the classical (unweighted) Lomb–Scargle method (Lomb 1976; Scargle 1982).

(5) '*Absolute magnitude*' (mag): the absolute magnitude in the *Hipparcos* band employing the *Parallax* described in item (10) and neglecting interstellar absorption.

(6) '*False-alarm probability*': the probability that the maximum peak in the Lomb–Scargle periodogram (Scargle 1982) was due to noise rather than the true signal, employing the beta distribution as indicated by Schwarzenberg-Czerny (1998). The computation assumed a number of independent frequencies equal to the number of frequencies tested divided by an oversampling factor (estimated by the largest value between 1 and the inverse of the product of the frequency spacing employed and the time series duration). While this attribute is sensitive to irregular light variations, it also depends on the S/N ratio of the time series and may thus include a significant survey-dependent contribution.

(7) '*QSO variations*': the reduced $\chi^2$ of the source variability with respect to a parametrized quasar variance model, denoted by $\chi^2_{QSO}/\nu$ in Butler & Bloom (2011). In general, the best-fitting scalings should depend on the measured brightness and corresponding filter, but for the *Hipparcos* data we followed Richards et al. (2011) and used parameter values corresponding to the SDSS *g* band at a fixed magnitude of 19.

(8) '*P2p scatter: folded/raw*': the point-to-point scatter of the time series folded with twice the recovered period (measured by the sum of squared magnitude differences between successive measurements in phase) divided by the same quantity computed on the raw time series (i.e. with respect to successive measurements in time). In the presence of periodicity, the scatter of successive points in folded light curves was expected to be smaller than the one related to time-sorted data (excluding periodicities of the order of the time series duration). This attribute achieved greater importance by folding the time series with *twice* the period because of the presence of eclipsing binaries. In fact, periods recovered with the Lomb–Scargle method were quite consistently half of the correct values for eclipsing binaries (Dubath et al. 2011). Thus, doubling the folding period reduced the point-to-point scatter, especially when the primary and secondary transit dips in the light curves of eclipsing binaries were different or not equally separated in phase.

(9) '*Scatter: raw/res*': the ratio between the median of absolute deviations from the median of the raw time series and the median of absolute values of the residual time series (obtained by subtracting model values from the raw time series), which provided a robust measure of the goodness of the model to represent the data.

(10) '*Parallax*' (mas): the parallax value from the new reduction of the *Hipparcos* raw data (van Leeuwen 2007a,b). Because of uncertainties involved in the measurement of parallax, about 10 per cent of the stars in the training set and 6 per cent in the prediction set had zero or negative parallax values. In order to use some of the information embedded in such measurements, non-positive values of parallax were replaced by positive values randomly extracted from a Gaussian distribution with zero mean and standard deviation equal to the measurement uncertainty.

(11) *Standard deviation res* (mag): the unbiased standard deviation of the residual time series, weighted by the inverse of squared measurement uncertainties.

(12) '*Short variations*' (mag): the average of absolute values of magnitude differences between all pairs of measurements separated by time-scales from 0.01 to 0.1 d.

(13) '*Sum squares: res/raw*': the ratio between the sum of squared residuals of the model from the raw data and the sum of squared deviations of the raw time series from its mean value.

(14) ' |*Galactic Latitude* | ' (deg): the absolute value of the Galactic latitude of the source position, possibly useful to distinguish younger stellar populations from older ones for the farthest objects.

(15) '*Frequency error*' (d$^{-1}$): the error estimate of the derived frequency under the assumption of equidistant observations of a sinusoidal signal (Kovács 1981; Baliunas et al. 1985; Gilliland & Fisher 1985).

The values of the attributes described above for sources in the training set are listed in Table 2, together with the *Hipparcos* identifiers and variability types from the literature. The distributions of training-set attributes are shown in Figs A1–A3 for each variability type (expressing some of the attributes in terms of decadic logarithms).

### 5.2 Period limitations

The periods recovered for classification with automated procedures are affected by the limitations described below and should be investigated further to confirm or amend their values for sources of interest.

First, time series were modelled with a single period, as deemed satisfactory in the classification of the *Hipparcos* periodic variables by Dubath et al. (2011). The focus of the present work on sources whose periods could not be determined accurately, or even confirmed with values from the literature, gave further reasons for not to pursue the search of multiple frequencies.

Secondly, as indicated by Dubath et al. (2011), periods of eclipsing binaries and ellipsoidal variables computed with the Lomb–Scargle method are most likely half of the correct values, as a consequence of the presence of two minima per cycle.

**Table 2.** The data refer to the training-set stars selected from the *Hipparcos* periodic, unsolved and microvariable sets. They include the *Hipparcos* identifiers, the values of the 15 most relevant attributes for classification and variability types from the literature (mostly from Watson et al. 2011). Limitations concerning the values of absolute magnitude, parallax and period are described in Section 5. This is only a portion of the full table available in the online version of the paper (see supporting information) and also from the Centre de Données astronomiques de Strasbourg website (http://cdsarc.u-strasbg.fr).

| Hip | $V-I$ | Skewness | Log(Amplitude) | Log(Period) | Absolute mag. | Log(False-alarm probability) | Log(P2p scatter: folded/raw) | Log(QSO var.) | Log(Scatter: raw/res) | Log(Parallax) | Log(Standard deviation res) | Log(Short var.) | Sum sq.: res/raw | \| Galactic Lat.\| | Frequency error $(\times 10^4)$ | Type |
|---|---|---|---|---|---|---|---|---|---|---|---|---|---|---|---|---|
| 8 | 3.92 | 0.61 | 0.61 | 2.5229 | 2.04 | −20.61 | −0.32 | 1.79 | 0.41 | 0.70 | −0.49 | −1.41 | 0.08 | 35.56 | 0.50 | *LPV_P* |
| 63 | −0.03 | −0.62 | −1.38 | 0.5727 | −0.02 | −28.66 | −0.15 | 0.00 | 0.30 | 0.73 | −2.21 | −2.13 | 0.19 | 16.70 | 0.22 | *ACV_P* |
| 99 | 2.91 | −0.04 | −0.46 | 0.6099 | −2.74 | −14.02 | 0.64 | 0.34 | 0.21 | 0.06 | −1.07 | −1.89 | 0.42 | 1.91 | 0.35 | *LPV_X* |
| 109 | 0.45 | −0.18 | −1.41 | −0.7819 | 1.13 | −2.88 | −0.06 | 0.64 | 0.00 | 0.76 | −1.82 | −1.70 | 0.69 | 53.98 | 0.57 | *DSCTC_P* |
| 139 | 1.25 | 1.34 | −1.23 | −0.3209 | 2.04 | −1.63 | 0.21 | 0.15 | −0.03 | 0.38 | −1.42 | −1.45 | 0.78 | 5.09 | 0.82 | *EA_X* |
| 154 | 2.35 | −0.10 | −0.98 | 1.1972 | −1.20 | −23.57 | 0.14 | 0.22 | 0.42 | 0.88 | −1.79 | −2.15 | 0.23 | 65.83 | 0.25 | *LPV_X* |
| 181 | 2.46 | 0.04 | −1.33 | −1.0367 | −1.01 | −7.61 | 0.41 | 0.06 | 0.09 | 0.44 | −1.66 | −1.95 | 0.66 | 1.60 | 0.57 | *LPV_X* |
| 226 | 0.29 | −0.75 | 0.10 | −0.3069 | 3.31 | −20.18 | −0.86 | 2.27 | 1.01 | 0.56 | −1.35 | −0.55 | 0.01 | 78.86 | 0.07 | *RRAB_P* |
| 270 | 0.16 | 2.00 | −0.71 | −0.1576 | 0.80 | −8.80 | −0.04 | 1.66 | −0.16 | 0.68 | −1.11 | −1.30 | 0.51 | 10.63 | 0.51 | *EA_P* |
| 274 | 0.43 | 0.32 | −1.41 | 0.4595 | −3.08 | −9.44 | 0.17 | −0.02 | 0.08 | 0.12 | −1.78 | −2.06 | 0.59 | 1.26 | 0.55 | *ACYG_X* |

Thirdly, spurious periods are expected to arise from aliased peaks in the periodogram (introduced by measurements sampled at discrete times) and parasite frequencies (due to features of the detection process). The *Hipparcos* satellite orbited the Earth with a period of 10 h and 42 min and scanned the sky with two distinct fields of view simultaneously. As a consequence of the spin of the satellite (with a period of 128 min) and the angle between the two fields, sources were typically observed in sequences of four to six transits (or many more for stars at the nodes of the scanning great circle) separated by 20 and 108 min, which repeated every three to five weeks (Eyer et al. 1994).

A set of spurious periods, among the ones expected from the above-mentioned time-scales, is easily recognized from the bimodal distribution of periods for some variability types (see Fig. A1, item 5): the narrow concentration of recovered periods around log $P \ll$ −1.05 corresponds to the rotational period of the satellite, indicating parasite frequencies possibly related to instrumental differences between the two fields of view. The variability types which seemed to be affected the most by this artefact included *I*, *LPV*, *RS+BY*, *BE+GCAS*, *ACV* and *EA* sources, especially the ones not selected from the *Hipparcos* periodic catalogue, which was processed carefully by the Geneva (Eyer 1998) and Cambridge (van Leeuwen 1997) teams. On the other hand, genuine periods of other types (such as *DSCT* stars) overlapped the same range too.

An automated procedure to remove aliased and parasite frequencies might be particularly challenging in the regime of low S/N ratios (e.g. see Reegen 2007), and it was not pursued in this work. The presence of sources with spurious periods in the training set might have reduced the classification power of the period values, but it helped recognize the variability types of sources associated with spurious periods in the prediction set. Furthermore, in the specific case of the parasite period at log $P \ll$ −1.05, the dependence on period to identify most of the affected types (listed in the previous paragraph) is negligible, as shown in Fig. 3.

Readers interested in obtaining more appropriate estimates of periods than those derived herein are suggested to consider the following.

(i) The search of multiple periods might improve the model in certain cases.

(ii) If the predicted type is *EA*, *EB*, *EW* or *ELL*, then the true period is most likely twice the quoted value. Extra caution is recommended for *EA*-type candidates, which might include faint outliers due to inaccurate instrumental pointing (Eyer & Grenon 2000).

(iii) Since the periodogram of a discretely sampled signal is the convolution of the periodogram of the continuous signal with the spectral window function, the latter may be employed to identify all major aliases (see Dawson & Fabrycky 2010, for a possible method).

# 6 RANDOM FORESTS

## 6.1 Method

Random forest is a tree-based classification method (Breiman 2001) which involves recursive binary partitioning of the attribute space. It is able to discover complex interactions among attributes and return more accurate classifiers than other methods employed in the literature, while demonstrating robustness to outliers, strongly correlated attributes and tuning parameters (Dubath et al. 2011; Richards et al. 2011).

For a training set of $N_{ts}$ objects, each characterized by $N_a$ attributes, the following items outline the basics of the method.

(1) A tree is built starting from a sample of $N_{ts}$ objects randomly drawn with replacement from the training set.
(2) A random subset of $m_{try}$ / $N_a$ attributes is selected to determine which one provides the best split at a given node of the tree.[13]
(3) The tree is fully developed until nodes include only single types.
(4) Typically, hundreds to thousands of decision trees are trained and they define the 'forest'.

---

[13] The tuning parameter $m_{try}$ is constant for all trees. Classification results are only weakly dependent on the exact choice of $m_{try}$, and the default value suggested for classification (the square root of the number of attributes; Liaw & Wiener 2002) has been employed herein.

(5) The predicted type of an object corresponds to the most frequent type predicted by the forest (each tree returns a single prediction per object).

The random draw with replacement, described in item (1), omits about one-third of the objects from the learning process for each tree. Such 'out-of-bag' objects are used to evaluate the misclassification rate of the tree they were left out from (and eventually of the overall classifier) by predicting their types and comparing them with the original ones.

This work employed the RANDOMFOREST package implemented in R (R Development Core Team 2011) by Liaw & Wiener (2002).

### 6.2 Attribute importance

The importance of a given attribute was assessed by the effect on the classification of shuffling its values among out-of-bag objects: the more important an attribute, the less accurate the classification of such objects becomes after rearranging its values randomly. Thus, the importance of an attribute was estimated by the mean increase in the misclassification rate after permuting the values of that attribute in the out-of-bag objects. The ranking of the most relevant attributes naturally depends on the composition of the training set (variability types and relative incidences). The most important attributes selected by Dubath et al. (2011) for periodic sources, such as colour, skewness, amplitude, absolute magnitude, period, and various scatter estimators and comparators, are among our topmost relevant attributes too, in addition to other attributes sensitive to stochastic light variations (indicating specific time-scales, affinity to QSO-like variations and level of confidence in the detected periodicity). The least important attributes in the top 15 list (e.g. from item 10 in Section 5.1) have marginal significance and are expected to depend on the survey.

The 15 most important attributes were employed to compute the average misclassification rate from 10-fold cross-validation experiments employing 2000 trees, following the procedure described by Dubath et al. (2011). The evolution of the average error rate as a function of the number of most important attributes is shown in Fig. 1. The first six most relevant attributes demonstrate the ability to drive the misclassification rate under 20 per cent, which is lowered to 16.6 per cent with a few additional attributes.[14] Aggregating sources from the periodic and unsolved/microvariable sets (labelled with suffixes 'P' and 'X', respectively) which refer to the same variability types, the lowest misclassification rate is under 12 per cent.

Due to the randomness intrinsic to random forests, the ranking of attributes might change after repeating the training of the classifier. The distribution of ranks of the 15 most important attributes is shown in Fig. 2 in terms of the percentage of occurrence of a certain rank for a given attribute.

The level of importance of each attribute for a particular run of random forest (with 2000 trees) is depicted for each variability type in Fig. 3 (where the darkness of the grey level of a cell in the matrix of attributes versus types is proportional to the attribute importance).

### 6.3 Confusion matrix

The classification results of the training set are shown in the confusion matrix in Fig. 4, with literature and predicted types listed in rows and columns, respectively. The overall misclassification rate of 16.3 per cent reduces to 11.6 per cent, waiving the additional information provided by the type label suffixes 'P' and 'X'. Detailed misclassification and contamination rates per variability type are presented in Table 3. The random forest classifier included 2000 trees and employed the 15 most important attributes described in Section 5.

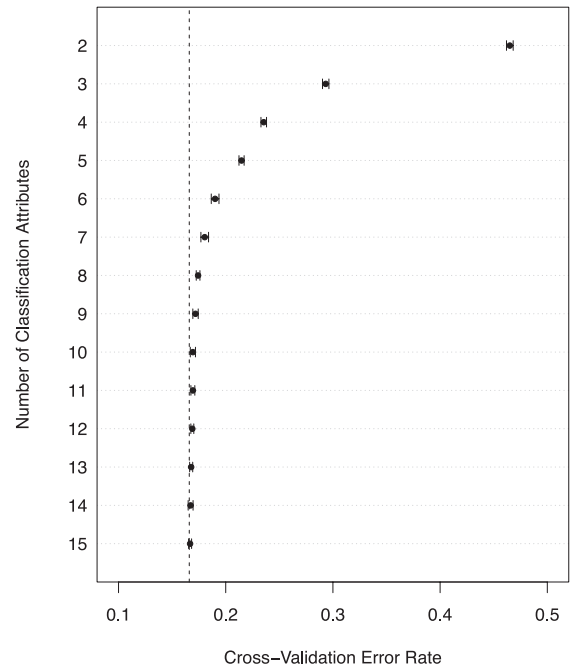

**Figure 1.** The evolution of the average misclassification rate from 10 runs of 10-fold cross-validation experiments employing 2000 trees in random forests is shown as a function of the number of the most important attributes. The first six most relevant attributes drive the error rate under 20 per cent, and a few more attributes lead to a misclassification rate of 16.6 per cent (dashed line), which corresponds to an overall error rate under 12 per cent if we aggregate sources that correspond to the same variability types in the periodic, unsolved and microvariable sets.

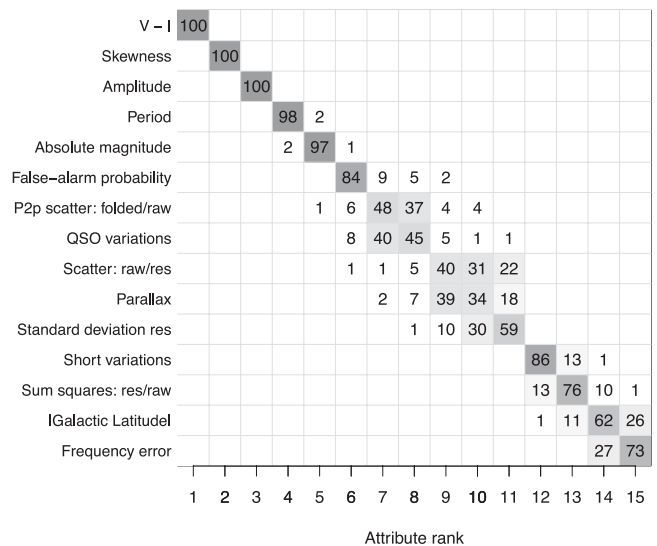

**Figure 2.** The distribution of ranks of the 15 most important attributes for classification is shown in terms of the percentage of occurrence of a certain rank for a given attribute. The table was produced by 10 runs of 10-fold cross-validation random forest experiments, and the importance of a given attribute was measured by the mean increase in the misclassification rate after permuting the values of that attribute in the out-of-bag sources. The Spearman correlation coefficient between any pair of attributes is smaller than 80 per cent.

---

[14] No improvement in the overall average accuracy of classification was achieved by employing the full set of 150 attributes.

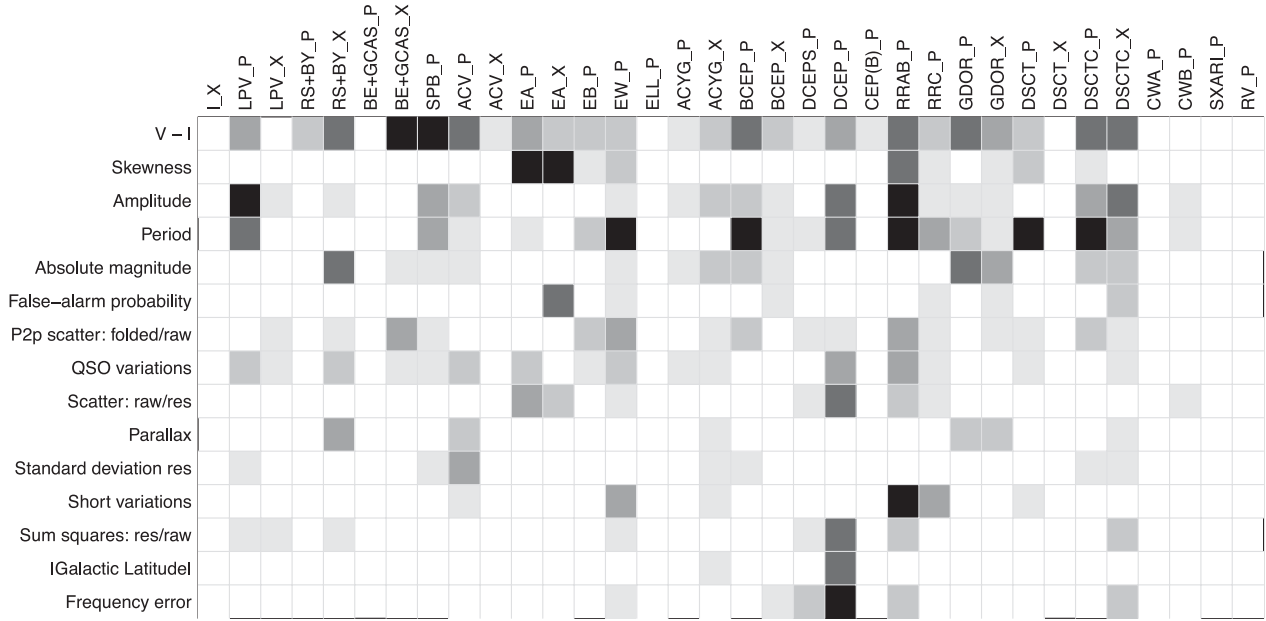

**Figure 3.** The importance of each attribute is depicted for each variability type, and it is proportional to the darkness of the grey level of the square corresponding to a given attribute-type combination. The importance of an attribute was measured in random forests by the mean increase in the misclassification rate after permuting the values of that attribute in the out-of-bag sources.

About one-third of the irregular variables was classified as semiregular (within the *LPV_X* type) and, to a lesser degree, as *RS+BY_X* and *BE+GCAS_X* types. This confusion is not surprising as the sources included in the *I* type are by definition not well studied and might actually belong to other types.

While the confusion between the slow irregulars and semiregulars merged in the *LPV_X* type was expected, the latter and its periodic counterpart (*LPV_P*) were distinguished quite clearly.

The *Hipparcos* unsolved catalogue provided several sources which helped characterize the *BE+GCAS* type: about 92 per cent of them were classified correctly, including over half of the periodic representatives, none of which was identified successfully in the companion paper devoted to the *Hipparcos* periodic variables (Dubath et al. 2011).

The *BE+GCAS_X* type was the most common prediction of misclassified variables of the following types (with an approximate percentage of confusion with *BE+GCAS_X* indicated in brackets, with respect to the number of sources of the true types).

(i) *BCEP_X* (50 per cent) and *ACYG_X* (26 per cent), possibly because of the presence of multiperiodicity (not considered in this work) or *B*-type stars with time series characterized by low S/N ratios. The lower confusion level of *ACYG_X* stars is probably a consequence of their greater brightness.

(ii) *ACV_X* (30 per cent), understood in terms of its location in the Hertzsprung–Russell diagram (similar to the *BE+GCAS_X* type) and common ranges of other attributes relevant for the identification of *ACV_X* stars.

(iii) *SPB* (11 per cent), possibly due to the late-type *BE* stars located inside the *SPB* instability strip in the Hertzsprung–Russell diagram.

The group of eclipsing binaries presented a level of confusion among the *EA*, *EB* and *EW* types consistent with the natural overlap between the *EA/EB* and *EB/EW* types intrinsic to their definition. Half of the *ELL* types were misclassified as eclipsing binaries, in agreement with the results of Dubath et al. (2011) and as expected from the similarities of the origins of these variability types.[15]

The confusion between the *CEP(B)* and *DCEP* types was expected because multimode pulsations could not be detected by modelling with a single period (as explained in Section 5.2, the search of multiple periods was not pursued herein).

One-quarter of *RRC* types were misclassified as *EW*-type eclipsing binaries, reflecting the similarities of the light-curve shapes (nearly symmetric, sometimes sinusoidal) and common ranges in the distributions of periods and other attributes.

About 94 per cent of the sources in the *DSCT* group (including SX Phoenicis and *DSCTC*) were identified correctly, with a significant distinction of the *DSCTC_X* types from their monoperiodic counterparts *DSCTC_P* (which had higher amplitudes, lower frequency errors and false-alarm probabilities, and were generally better modelled).

Finally, the *CWA*, *CWB*, *SXARI* and *RV* types did not benefit of additional training-set sources from the *Hipparcos* unsolved catalogue or a set of microvariables, remaining poorly represented and with little or no success of identification, as found by Dubath et al.

---

[15] Similar to eclipsing binaries, ellipsoidal rotating variables originate from binary systems viewed nearly edge-on. When the binary stars are close enough to distort their shapes (e.g. into ellipsoids), their orbital motion can cause variations in the combined visible surface area of the stars, so the overall brightness can vary even in the absence of eclipses.

| | I_X | LPV_P | LPV_X | RS+BY_P | RS+BY_X | BE+GCAS_P | BE+GCAS_X | SPB_P | ACV_P | ACV_X | EA_P | EA_X | EB_P | EW_P | ELL_P | ACYG_P | ACYG_X | BCEP_P | BCEP_X | DCEPS_P | DCEP_P | CEP(B)_P | RRAB_P | RRC_P | GDOR_P | GDOR_X | DSCT_P | DSCT_X | DSCTC_P | DSCTC_X | CWA_P | CWB_P | SXARI_P | RV_P |
|---|---|---|---|---|---|---|---|---|---|---|---|---|---|---|---|---|---|---|---|---|---|---|---|---|---|---|---|---|---|---|---|---|---|---|
| I_X | 15 | | 18 | | 5 | | 8 | | 1 | | 1 | | | | | 1 | | 1 | | | | | | | 2 | 1 | | | | | | | | |
| LPV_P | | 250 | 35 | | | | | | | | | | | | | | | | | | | | | | | | | | | | | | | |
| LPV_X | | 15 | 1437 | 6 | | 5 | | | | | | | | | | 1 | | | | | | | | | | | | | | | | | | |
| RS+BY_P | | 3 | 8 | 23 | | | | | | | | | | | | | | | | | 1 | | | | | | | | | | | | | |
| RS+BY_X | 2 | | 8 | 7 | 120 | | | | | | 1 | | | | | | | | | | | | | | | | | | | | | | | |
| BE+GCAS_P | | | | | | | | 7 | 2 | 2 | 1 | | | | | 1 | | | | | | | | | | | | | | | | | | |
| BE+GCAS_X | 1 | | | | | | 235 | 2 | 1 | 3 | | | 2 | | | 3 | | 2 | | | | | | | | | | | | | | | | |
| SPB_P | | | | | | 9 | | 64 | 5 | | 1 | | | | 1 | | | | | | | | | | | | | | | | | | | |
| ACV_P | | | | 1 | | | | 7 | 63 | 2 | | | 3 | | | | | | | | | | | | | | | | | | | | 1 | |
| ACV_X | | | | 1 | | | | 8 | 2 | 7 | 4 | | | | | | | | | | | | | | | | | | | | | | 3 | |
| EA_P | | | 1 | | | | | | | 2 | 179 | 25 | 21 | | | | | | | | | | | | | | | | | | | | | |
| EA_X | | | 2 | | | | | 3 | | | 23 | 101 | | | | | | | | | | | 1 | | | | | | | | | | | |
| EB_P | | | 3 | | 1 | | | 4 | 7 | 2 | 26 | 2 | 176 | 24 | 2 | 1 | | | | 1 | | 1 | | | | 1 | | | | 4 | | | | |
| EW_P | | | | | | | | | | | | | 29 | 77 | | | | | | | 1 | | | | | | | | | | | | | |
| ELL_P | | | 2 | | 1 | | | | 1 | | 1 | | 4 | | 14 | | | | | | | 1 | | | | | | | | | | | | |
| ACYG_P | | | 1 | | | | | 3 | 1 | | | | | | 2 | 11 | | | | | | | | | | | | | | | | | | |
| ACYG_X | | | 2 | | | | | 9 | | | | | | | | 2 | 21 | | | | | | | | | | | | | | | | | |
| BCEP_P | | | | | | | | | 1 | 2 | | | 2 | | | | | 22 | 1 | | | | | | | | | | 2 | | | | | |
| BCEP_X | | | | | | | | 13 | | | | | | | | | 1 | 3 | 9 | | | | | | | | | | | | | | | |
| DCEPS_P | | | 1 | | | | | | | | | | 1 | | | | | | | 14 | 12 | 1 | | | | | | | | | | | | |
| DCEP_P | | 1 | | | | | | | | | | | | | | | | | | 8 | 179 | 1 | | | | | | | | | | | | |
| CEP(B)_P | 1 | | 1 | | | | | | | | | | | | | | | | | | 2 | 5 | 1 | | | | | | | | | | | |
| RRAB_P | | | | | | | | | | | | | 3 | | | | | | | | | | 68 | 1 | | | | | | | | | | |
| RRC_P | | | | | | | | | | | | | 2 | 5 | | | | | | | | | 1 | 11 | | | | 1 | | | | | | |
| GDOR_P | | | | | | | | | | | | | | | | | | | | | | | | | 23 | 4 | | | | | | | | |
| GDOR_X | | | | | | | | | | | | | | | | | | | | | | | | | 6 | 9 | | 2 | | | | | | |
| DSCT_P | | | | | | | | | | | | | 2 | 1 | | | | | | | 1 | 1 | | | | | 31 | 8 | 3 | | | | | |
| DSCT_X | 2 | | | | | | | 1 | | | 1 | | | | | | | | | | | | | | | | 6 | 10 | | | | | | |
| DSCTC_P | | | | | | | | | | | | | 1 | | | | | | | | | | | 1 | | | | 1 | 71 | 7 | | | | |
| DSCTC_X | | | | | | | | | | | | | 1 | | | | | | | | | | | 1 | | | | 1 | 6 | 52 | | | | |
| CWA_P | 1 | | 1 | | | | | | | | | | | | | | | | | | | 4 | | | | | | | | | 1 | | | |
| CWB_P | | | | | | | | | | | | | 1 | | | | | | | | 2 | 3 | | | | | | | | | | | | |
| SXARI_P | | | | | | | | 3 | 3 | | | | | | | | | | | | | | | | | | | | | | | | | |
| RV_P | | | 1 | | | | | | 1 | | | | 1 | | | | | | | | | 2 | | | | | | | | | | | | |

**Figure 4.** The confusion matrix is shown for the training-set sources classified by random forests with 2000 trees and employing the 15 most important attributes. Labels of variability types are described in Table 1, with suffixes 'P' and 'X' indicating the *Hipparcos* catalogue from which sources have been selected (periodic and unsolved/microvariable, respectively). Rows and columns refer to literature and predicted types, respectively. The overall misclassification rate of 16.3 per cent reduces to 11.6 per cent, waiving the additional information provided by the label suffixes 'P' and 'X'.

(2011). The confusion of almost half of the W Virginis and RV Tauri types (subclasses of Population II Cepheids) with classical Cepheids (stars of Population I) was expected as a result of uncertain absolute magnitudes and lack of metallicity information.

### 6.4 Impact of imbalanced data

The numbers of representatives of variability types in the training set (listed in Table 1) differ by up to three orders of magnitude. Since the classifier aims at minimizing the overall misclassification rate, the most populated classes might bias the classification. This behaviour would not be a disadvantage if the relative incidence of sources per type in the training set matches the one in the prediction set, and if the highest number of correct classifications is pursued (penalizing rare types). The training set employed in this work is clearly dominated by the *LPV* types. In order to test the potential deterioration of the classification of other variability types in the presence of such a dominant type, *LPV* and non-*LPV* candidates were first separated by a dedicated classifier (with its own relevant attributes), and successively the two groups were classified into all of the variability types they contained, each employing different classifiers and sets of attributes. After aggregating the classification results related to the same types, the classification accuracies of the single- and two-stage approaches were comparable within noise fluctuations for all types. Such a result was likely caused by the low error rate associated with the *LPV* type, with false positive and false negative rates at the level of 1 per cent. A few attributes, such as colour, amplitude and period, were sufficient to identify most of the *LPV* types, so a relevant effort in the minimization of the overall misclassification rate could be devoted to the other variability types. A similar conclusion was also applied to the second most numerous group of types (eclipsing binaries and ellipsoidal variables), so the most populated variability types were not deemed detrimental to the classification of other types.

**Table 3.** Misclassification and contamination per cent rates derived from the confusion matrices obtained with random forests (RF) and the multistage Bayesian networks (MB), presented in Figs 4 and 5, respectively, aggregating label suffixes 'P' and 'X' referring to the same variability types. The numbers in bold refer to sets of related types which may be legitimately confused.

| Type | Misclassification | | Contamination | |
|---|---|---|---|---|
| | RF | MB | RF | MB |
| *I* | 72.2 | 61.1 | 31.8 | 51.2 |
| *LPV* | 0.7 | 0.7 | 2.5 | 2.6 |
| *RS+BY* | 8.7 | 12.7 | 10.2 | 11.7 |
| *BE+GCAS* | 7.6 | 8.0 | 21.9 | 17.5 |
| *SPB* | 21.0 | 21.0 | 29.7 | 34.0 |
| *ACV* | 26.9 | 23.1 | 24.0 | 26.6 |
| ***Eclipsing & ELL*** | **6.4** | **7.2** | **4.5** | **4.7** |
| *EA* | 8.6 | 7.2 | 10.1 | 5.9 |
| *EB* | 31.0 | 29.4 | 32.0 | 27.4 |
| *EW* | 28.0 | 26.2 | 28.0 | 31.9 |
| *ELL* | 100.0 | 96.3 | 100.0 | 90.0 |
| *ACYG* | 32.1 | 32.1 | 20.0 | 14.3 |
| *BCEP* | 37.5 | 30.4 | 16.7 | 15.2 |
| ***Delta Cephei (I & II)*** | **7.2** | **4.8** | **1.7** | **2.1** |
| *DCEPS* | 54.8 | 35.5 | 39.1 | 31.0 |
| *DCEP* | 5.3 | 4.2 | 12.3 | 10.0 |
| *CEP(B)* | 54.5 | 63.6 | 28.6 | 33.3 |
| *CWA* | 85.7 | 100.0 | 0.0 | 100.0 |
| *CWB* | 100.0 | 50.0 | (n/a) | 25.0 |
| *RV* | 100.0 | 100.0 | (n/a) | 100.0 |
| ***RR Lyrae*** | **12.0** | **10.9** | **8.0** | **10.9** |
| *RRAB* | 5.6 | 5.6 | 9.3 | 6.8 |
| *RRC* | 45.0 | 35.0 | 15.4 | 31.6 |
| *GDOR* | 4.5 | 29.5 | 10.6 | 6.1 |
| ***Delta Scuti*** | **6.2** | **8.6** | **7.1** | **6.4** |
| *DSCT* | 53.7 | 59.7 | 11.4 | 20.6 |
| *DSCTC* | 4.2 | 7.7 | 22.7 | 22.9 |
| *SXARI* | 100.0 | 100.0 | (n/a) | (n/a) |

## 7 MULTISTAGE BAYESIAN NETWORKS

The study of a multistage classifier that incorporates Bayesian networks at the nodes of a classification tree is presented for comparison. This classification model (or evolutions of it) has been used as a standard for comparison since it was first proposed for the classification of variable stars by Debosscher et al. (2007) and Sarro et al. (2009). The multistage classifier is based on a divide-and-conquer approach whereby the classification task in the framework of the complete set of 34 variability types (without aggregating label suffixes referring to the same types) is separated into smaller and simpler subtasks of dichotomic classification into two categories, each of which can be a compound of several variability types. The multistage classification can be represented by a tree with connecting nodes that represent each dichotomic classifier (see Fig. B1 for the actual tree used in this work). More details about the algorithm of the multistage scheme definition are given below and can also be found in Dubath et al. (2011). Herein, instead of a single classifier at each node, the so-called bagging approach was used (Breiman 1996), which is equivalent to the bootstrapping of classification trees used in the construction of a random forest.

The bagging approach was introduced in order to overcome the large increase in the misclassification rate of the classifier based on simple Bayesian networks. While the previous versions of the classifier described by Dubath et al. (2011) yielded equivalent misclassification rates for the random forest and multistage classifiers (at the 15.7 per cent level), for the expanded data set presented herein the multistage classifier with Bayesian networks in the nodes produced a 19.7 per cent misclassification rate compared to the 16.3 per cent achieved by random forests (see Fig. 4). This discrepancy was interpreted as a result of the inclusion of cases with S/N ratios significantly lower than those characterizing time series in the *Hipparcos* catalogue of periodic variables. In this context, the bagging approach was expected to minimize the impact of these reduced S/N ratios by making the process of attribute discretization fuzzier. The fuzzification comes as a result of having different discretizations for each bootstrap sample. One thousand replicates were used in each node, and each replicate was constructed from a bootstrap sample with the same size as the training set.

The bagging approach was also used for the monolithic classifier that provided the basic information to structure the multistage tree (see Dubath et al. 2011, and Fig. B1). The algorithm used to define the multistage tree is based on the confusion matrix obtained using a monolithic, single-stage classifier. From this, the similarity $S$ between classes is determined according to the metrics

$$S(T_i, T_j) = \begin{cases} 1 & \text{if } i = j, \\ \frac{X_{ij} + X_{ji}}{X_{ii} + X_{jj} + X_{ij} + X_{ji}} & \text{if } i \cong j, \end{cases} \tag{1}$$

where $T_i$ represents type $i$ and $X_{ij}$ represents the element in the $i$th row and $j$th column of the confusion matrix. The idea behind this metric is to try to isolate groups of types with small intergroup contamination rates and (in some cases) significant intragroup overlap. Types that are easily separable should appear in the topmost levels of the tree to simplify the classification of other types and select optimal sets of attributes for the tree nodes that correspond to types with significant overlaps. The tree-definition algorithm starts from the complete set of types and merges, in each step, the two most similar types into a new type. The similarities are then recalculated over the new set of types, and a new merge is proposed according to them. The history of merged types is then reversed to define the multistage tree.

The 10-run 10-fold cross-validation experiment (see Bouckaert 2003) yielded an average misclassification rate of $16.63 \pm 0.18$ per cent (corresponding to an error rate of 11.38 per cent when label suffixes for the same types were aggregated).

The corresponding confusion matrix is shown in Fig. 5, and this classifier model has been applied to predict new variability types (as described in Section 8). These experiments were carried out starting with the full set of 150 attributes originally created for the *Hipparcos* sources and then selecting the optimal set of attributes for each node as described in Dubath et al. (2011).

In order to facilitate the comparison of results obtained with the multistage Bayesian networks and random forests, detailed misclassification and contamination rates per type are presented in Table 3 for both methods (employing the confusion matrices in Figs 4 and 5). Variability types which are related and might be legitimately confused have been considered separately as well as grouped together. While the overall accuracies of the two classifiers are similar, there are significant type-dependent variations. Such results provide important indications for the interpretation of type predictions when these are generated by both methods for the same sources.

Figure 5 — confusion matrix:

| | I_X | LPV_P | LPV_X | RS+BY_P | RS+BY_X | BE+GCAS_P | BE+GCAS_X | SPB_P | ACV_P | ACV_X | EA_P | EA_X | EB_P | EW_P | ELL_P | ACYG_P | ACYG_X | BCEP_P | BCEP_X | DCEPS_P | DCEP_P | CEP(B)_P | RRAB_P | RRC_P | GDOR_P | GDOR_X | DSCT_P | DSCT_X | DSCTC_P | DSCTC_X | CWA_P | CWB_P | SXARI_P | RV_P |
|---|---|---|---|---|---|---|---|---|---|---|---|---|---|---|---|---|---|---|---|---|---|---|---|---|---|---|---|---|---|---|---|---|---|---|
| I_X | 21 | | 20 | 1 | | 6 | | | 1 | | | | | | | 1 | 1 | | | | 1 | | | | | | | | | | 1 | | | |
| LPV_P | | 253 | 32 | | | | | | | | | | | | | | | | | | | | | | | | | | | | | | | |
| LPV_X | 2 | 16 | 1435 | 1 | 5 | 5 | | | | | | | | | | | | | | | | | | | | | | | | | | | | |
| RS+BY_P | | | | 5 | 9 | 21 | | | | | | | | | | | | | | | | | | | | | | | | | | | | |
| RS+BY_X | 8 | | | 6 | 9 | 112 | | | | | | | | | | | | | | | | | | | | | | | | | | | | |
| BE+GCAS_P | | | | | | 7 | 3 | 1 | | | | 1 | | | | 1 | | | | | | | | | | | | | | | | | | |
| BE+GCAS_X | 2 | | | 1 | | 1 | 234 | 1 | 3 | | 1 | 1 | | | | 2 | | 3 | | | | | | | | | | | | | | | | |
| SPB_P | | | | | | 5 | | 64 | 12 | | | | | | | | | | | | | | | | | | | | | | | | | |
| ACV_P | 1 | | | | | | | 7 | 63 | 1 | | | 1 | | | | 2 | | | | | | | | | | | | | | | | 1 | |
| ACV_X | 5 | | | | | | | 2 | 10 | 6 | | | 1 | | | 1 | | | | | | | | | | | 1 | | | | | | 1 | |
| EA_P | | 1 | 1 | | | 2 | | | | | 182 | 30 | 12 | | | | | | | | | | | | | | | | | | | | | |
| EA_X | | 1 | 1 | | | 2 | | | 2 | 1 | 26 | 95 | 4 | | | | | | | | | | | | | | | | | | | | | |
| EB_P | | | 4 | 1 | 1 | 3 | 9 | | 1 | | 11 | 3 | 180 | 32 | 2 | 1 | | 1 | 1 | | | | | | | | | 4 | | | | | | |
| EW_P | | | | | | | | | | | | | 26 | 79 | | | | 1 | | | | | | | | | | 1 | | | | | | |
| ELL_P | | | 2 | 1 | | 1 | | | | | | | 4 | 4 | 11 | | | | | | | | | | | | 1 | 1 | | | | | | |
| ACYG_P | | 1 | 2 | | | 2 | | | | | | | | | | 3 | 10 | | | | | | | | | | | | | | | | | |
| ACYG_X | | 2 | 2 | | | 8 | | | | | | | | | | 5 | 18 | | | | | | | | | | | | | | | | | |
| BCEP_P | | | | | | | | 1 | | | | | | | 4 | | | 22 | 3 | | | | | | | | | | | | | | | |
| BCEP_X | | | | | | | | 5 | 4 | 2 | | | | | 1 | | | 2 | 12 | | | | | | | | | | | | | | | |
| DCEPS_P | 1 | | | 1 | | | | | | | | | | | 1 | | | 1 | | 20 | 6 | 1 | | | | | | | | | | | | |
| DCEP_P | 1 | | | | | | | | | | | | | | | | | | | 6 | 181 | 1 | | | | | | | | | | | | |
| CEP(B)_P | 2 | | | | | | | | | | | | | | | | | | | 4 | 4 | 1 | | | | | | | | | | | | |
| RRAB_P | | | | | | | | | | | | | | | 1 | 1 | | | | | | | 68 | 1 | | | | | | | | | 1 | |
| RRC_P | | | | | | | | | | | 1 | | 4 | | | | | 1 | | | | | | 13 | | | | 1 | | | | | | |
| GDOR_P | 1 | | | | | 2 | | | | | 1 | | | | 3 | | | | | | | | | | 18 | 1 | | 1 | | | | | | |
| GDOR_X | 1 | | | | | 1 | | | | | | | | | 1 | | | | | | | | | | 7 | 5 | | 2 | | | | | | |
| DSCT_P | | | | | | | | | | | | | | 2 | 1 | | | | | | | | | | 3 | 5 | 26 | | 7 | 3 | | | | |
| DSCT_X | 2 | | | | | | | | | | | | | | 1 | | | | | | | | | | | | 6 | 11 | | | | | | |
| DSCTC_P | | | | | | | | 1 | | | | | | | 1 | | | | | | | | | 1 | | | 4 | | 65 | 9 | | | | |
| DSCTC_X | 1 | | | | | | | | | | | | | | 1 | | | | | | | | | | | | 2 | | 8 | 49 | | | | |
| CWA_P | | | 1 | | | | | | | | | | | | | | | | | | 5 | | | | | | | | | | | | | 1 |
| CWB_P | | | | | | | | | | | | | | | | | | | | | 1 | 2 | | | | | | | | | | 3 | | |
| SXARI_P | | | | | | 1 | | | 2 | 2 | 1 | | | | 1 | | | | | | | | | | | | | | | | | | | |
| RV_P | | 2 | | | | | | | | | | | | | 1 | | | | | | | 2 | | | | | | | | | | | | |

**Figure 5.** The confusion matrix is shown for the training-set sources classified by a multistage classifier employing Bayesian networks and composed of bagging ensembles of dichotomic nodes, with 1000 replicates and feature selection based on mutual information at each node. The full set of about 150 attributes (prior to selection) is used by the classifier. Labels of variability types are described in Table 1, with suffixes 'P' and 'X' indicating the *Hipparcos* catalogue from which sources have been selected (periodic and unsolved/microvariable, respectively). Rows and columns refer to literature and predicted types, respectively. The misclassification rate is 16.5 per cent, which corresponds to an overall error rate of 11.4 per cent if label suffixes 'P' and 'X' referring to the same variability types are aggregated.

# 8 PREDICTIONS OF VARIABILITY TYPES

Predictions of variability types are provided for 6051 stars selected from the *Hipparcos* unsolved catalogue and a set of microvariables.

## 8.1 Prediction set

Types were predicted for only those sources whose current classifications were still missing or likely to be improved by our classifier models. Such cases included sources associated with types labelled as alternative, uncertain (but not combined), generic (such as periodic, pulsating, double, pre-main-sequence, miscellaneous or variable), unstudied (with slow or rapid light variations) and excluded secure types listed in the training set, secure or uncertain classifications of rare types (absent from the training set because represented by less than 12 objects), secure or suspected combinations of different types, hastily designated variables (not confirmed by further observations) and unique variabilities not consistent with known types (like brief transitions between types or future new types).

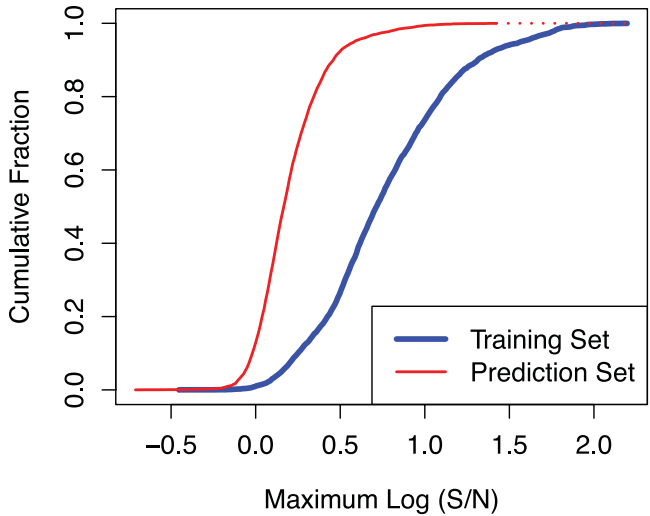

**Figure 6.** The cumulative fractions of sources from the training and prediction sets are presented as a function of the S/N ratio estimated by $\sqrt{(\sigma/\epsilon)^2 - 1}$, where $\sigma$ and $\epsilon$ denote the standard deviation and median error of a time series, respectively.

This work was limited to predictions of only single types (corresponding to the most likely type for each source), because patterns indicating possible combinations of different variability types were believed to arise mostly from the inaccuracy of the classifier or intrinsic confusion between types, in a regime of data characterized by irregular features and low S/N ratios (see Fig. 6).

In order to avoid the combination of non-ideal data quality and low number of measurements or incomplete source characterization, 10 sources with less than 30 observations were removed from the prediction set, and sources with missing attributes were excluded (which affected 158 objects, with missing information typically related to the colour, i.e. the most important attribute). Only data with good quality flags [field HT4 of ESA (1997) not set, or only bit 0 set] were considered.

The full list of sources for which variability types are predicted and the corresponding attribute values are listed in Table 4.

## 8.2 Results

Classification results are presented in Table 5, which includes variability types predicted by both random forests (Section 6) and the multistage methodology based on Bayesian networks (Section 7). Types from the literature (ESA 1997; Watson et al. 2011) are provided when available. Predictions are complemented by additional information to assess their reliability, such as the type probability and the *Hipparcos* sets from which the sources have been selected ($U_1$, $U_2$ and M, as defined in Section 3). The full probability arrays of classification of each source into all types are available only online from the Centre de Données astronomiques de Strasbourg website[16] as described in Appendix C. The numbers of predictions for each of such sets are 3070, 2037 and 944, respectively. Because of the low S/N ratio of most sources, all predictions should be interpreted with caution as general indicators of the variability type membership to be confirmed by further investigations.

The amount of useful predictions depends on the desired levels of completeness and purity. In general, the most reliable candidates are expected to occur in the following circumstances.

(i) High probability of the predicted types.
(ii) Same or related types predicted by both classifiers.
(iii) Sources selected from the *Hipparcos* set $U_1$.[17]
(iv) Predictions in agreement with literature information, when available.

Predictions are limited to the variability types included in the training set. Low-probability classifications with unexpected attribute values (such as the negative skewness of the magnitude distribution of eclipsing binary candidates) might be due to weak features or missing types in the training set.

Visual inspections of the light curves of *EA*-type candidates could not confirm the classifier predictions. Alternative period-search methods were tested too, but they did not improve the folded light-curve shapes sufficiently. Employing a subset of 156 *EA* candidates with the same type predicted by both classifiers, the rate of faint outliers (fainter than the 85th percentile of the magnitude distribution of each time series) was compared with the corresponding value related to brighter measurements as a function of survey time. The results are shown in Fig. 7, and they indicate an excess of faint outliers at the end of mission among data flagged as reliable. Such an effect might suggest the presence of occasional dimming induced by inaccurate instrumental pointing (ESA 1997; Eyer & Grenon 2000).

For the specific case of *GDOR* types, we compared our predictions with the results of Handler (1999), which included candidates from the *Hipparcos* unsolved catalogue too. Handler (1999) separated the most reliable *GDOR* candidates from the uncertain ones and listed them in tables 1 and 2, respectively. All of the 'prime' candidates from table 1, which are included in our prediction set,[18] are confirmed by the predictions of random forests and/or multistage Bayesian networks, and 15 further candidates[19] from table 2 (among the 25 stars in common with our prediction set) are also confirmed by at least one of the classifiers.

The probabilities $p_{RF}$ and $p_{MB}$ of variability types predicted by random forests and multistage Bayesian networks, respectively, are compared in Fig. 8. About 74 per cent of the sources (4428 objects) are associated with the same aggregated types (regardless of the 'P' and 'X' suffixes) by both classifiers, as indicated by the blue circles, typically at high-probability levels, while different types are predicted for the same source in about 26 per cent of the cases (1603 objects, denoted by red triangles and generally at low probabilities). For reference purposes, the dashed lines in Fig. 8 mark symmetric offsets with respect to the diagonal ($p_{RF} = p_{MB}$), defined by

---

**Table 4.** The data refer to the set of stars for which variability types are predicted, selected from the *Hipparcos* unsolved catalogue and a set of microvariables. They include the *Hipparcos* identifiers and the values of the 15 most relevant attributes for classification. Limitations concerning the values of absolute magnitude, parallax and period are described in Section 5. This is only a portion of the full table available in the online version of the paper (see supporting information) and also from the Centre de Données astronomiques de Strasbourg website (http://cdsarc.u-strasbg.fr).

| Hip | $V-I$ | Skewness | Log(Amplitude) | Log(Period) | Absolute mag. | Log(False-alarm probability) | Log(P2p scatter: folded/raw) | Log(QSO var.) | Log(Scatter: raw/res) | Log(Parallax) | Log(Standard deviation res) | Log(Short var.) | Sum sq.: res/raw | \|Galactic Lat.\| | Frequency error ($\times 10^4$) |
|---|---|---|---|---|---|---|---|---|---|---|---|---|---|---|---|
| 40 | 0.64 | 2.16 | −1.01 | −0.4819 | 2.92 | −0.62 | 0.10 | 0.59 | −0.08 | 0.52 | −1.16 | −1.16 | 0.81 | 4.83 | 1.11 |
| 57 | 0.88 | 0.04 | −1.51 | −1.0821 | 6.03 | −4.34 | 0.25 | 0.06 | 0.01 | 1.52 | −1.69 | −1.70 | 0.84 | 46.82 | 0.86 |
| 76 | 2.29 | 0.04 | −1.28 | 1.3072 | 1.40 | −4.84 | −0.05 | 0.08 | 0.19 | 0.46 | −1.67 | −1.72 | 0.66 | 28.86 | 0.62 |
| 107 | 1.88 | 0.54 | −1.64 | −1.0412 | −0.48 | −7.99 | 0.13 | 0.02 | 0.07 | 0.78 | −1.96 | −2.12 | 0.66 | 64.90 | 0.54 |
| 120 | 1.14 | −0.13 | −1.69 | −1.4739 | 0.38 | −1.56 | −0.06 | 0.10 | 0.13 | 0.38 | −1.79 | −1.63 | 0.88 | 69.25 | 1.10 |
| 174 | 1.01 | 0.13 | −1.51 | −0.2290 | 2.39 | −2.92 | 0.02 | 0.00 | 0.05 | 0.67 | −1.68 | −1.60 | 0.79 | 2.32 | 0.80 |
| 178 | 0.63 | 2.32 | −1.57 | −1.3518 | 2.91 | −0.13 | −0.14 | 0.48 | −0.02 | 0.96 | −1.65 | −1.65 | 0.85 | 79.07 | 1.18 |
| 215 | 2.37 | −0.29 | −1.07 | 1.7077 | 0.99 | −8.69 | −0.21 | −0.04 | 0.18 | 0.34 | −1.70 | −1.73 | 0.40 | 12.47 | 0.49 |
| 231 | 0.57 | −0.46 | −1.33 | −1.1200 | 2.43 | −1.32 | −0.06 | 0.14 | 0.07 | 0.55 | −1.56 | −1.40 | 0.75 | 57.24 | 1.10 |
| 235 | 1.18 | −0.37 | −1.23 | 0.0947 | 1.99 | −3.63 | −0.03 | 0.11 | 0.06 | 0.53 | −1.51 | −1.51 | 0.80 | 8.86 | 0.77 |

**Table 5.** Variability types predicted by random forests (RF) and from a multistage method based on Bayesian networks (MB) are listed together with the *Hipparcos* identifiers, the *Hipparcos* sets from which the sources have been selected (U₁, U₂ and M, as defined in Section 3), and literature variability types from the *Hipparcos* and the AAVSO catalogues when available. Each type prediction is associated with a probability (from RF and MB). This is only a portion of the full table available in the online version of the paper (see supporting information) and also from the Centre de Données astronomiques de Strasbourg website (http://cdsarc.u-strasbg.fr/). The latter also hosts the full probability arrays of classification of each source into all variability types (in the online Tables C1 and C2, see also the supporting information).

| Hip | Hipparcos set | Hipparcos type | AAVSO type | Predicted type RF | MB | Probability RF | MB |
|---|---|---|---|---|---|---|---|
| 40 | U₁ | – | E: | EA_X | EA_X | 0.66 | 0.66 |
| 57 | U₂ | – | – | RS+BY_X | RS+BY_X | 0.73 | 0.75 |
| 76 | U₁ | – | LB: | LPV_X | LPV_X | 0.79 | 0.93 |
| 107 | U₁ | – | – | LPV_X | LPV_X | 0.87 | 0.95 |
| 120 | U₂ | – | – | DSCTC_X | RS+BY_X | 0.30 | 0.25 |
| 174 | U₂ | – | – | RS+BY_X | RS+BY_X | 0.43 | 0.55 |
| 178 | U₁ | – | – | EA_X | EA_X | 0.64 | 0.76 |
| 215 | U₁ | – | LB: | LPV_X | LPV_X | 0.88 | 0.93 |
| 231 | U₂ | – | – | DSCT_X | DSCT_X | 0.21 | 0.52 |
| 235 | U₂ | – | – | RS+BY_X | RS+BY_X | 0.25 | 0.53 |

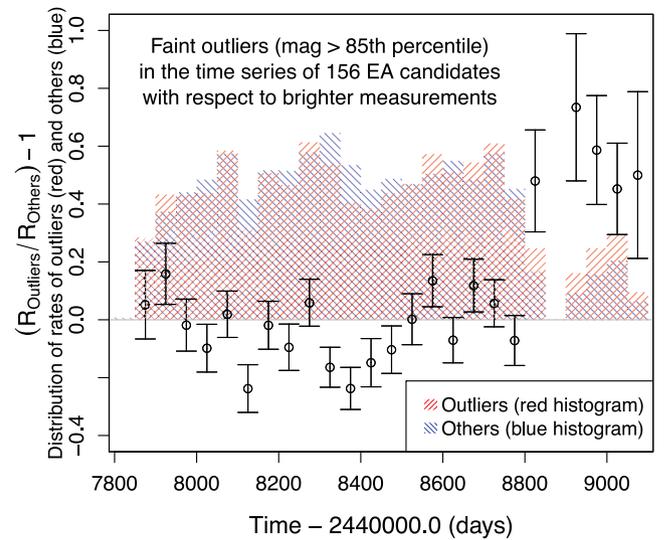

**Figure 7.** The ratio between the rate of faint outliers $R_{Outliers}$ (fainter than the 85th percentile of the magnitude distribution of each time series) and the rate of brighter measurements $R_{Others}$ is computed for a set of 156 *EA* candidates as a function of survey time (JD−244 0000.0). Each rate is normalized by the total number of outliers and non-outliers, respectively. The underlying histograms illustrate the rates of outliers and those of the other data points (multiplied by a factor of 10 for better visibility). Only measurements with good quality flags are included.

$|p_{RF} − p_{MB}| = 0.7\,(1 − |p_{RF} + p_{MB} − 1|)$, which help notice that multistage Bayesian networks tend to be more confident than random forests, on average.

The distribution of the probabilities of variability types predicted by random forests is shown in Fig. 9 as a function of the S/N ratio, and it shows that higher levels of probability and S/N ratio are populated mostly by the *LPV* types. The distribution of predictions per variability type is presented in Table 6 for selected thresholds of the predicted type probability. The Hertzsprung–Russell diagram[20] of stars with the same variability types predicted by both random forests and the multistage methodology involving Bayesian networks (with probabilities of 0.4 or greater according to both classifiers) is illustrated in Fig. 10, excluding eclipsing binaries and irregular variables because they are not associated with specific colour–magnitude loci by definition.

[20] Limitations apply to the values of absolute magnitude when parallax measurements are insignificant, as described in Section 5.

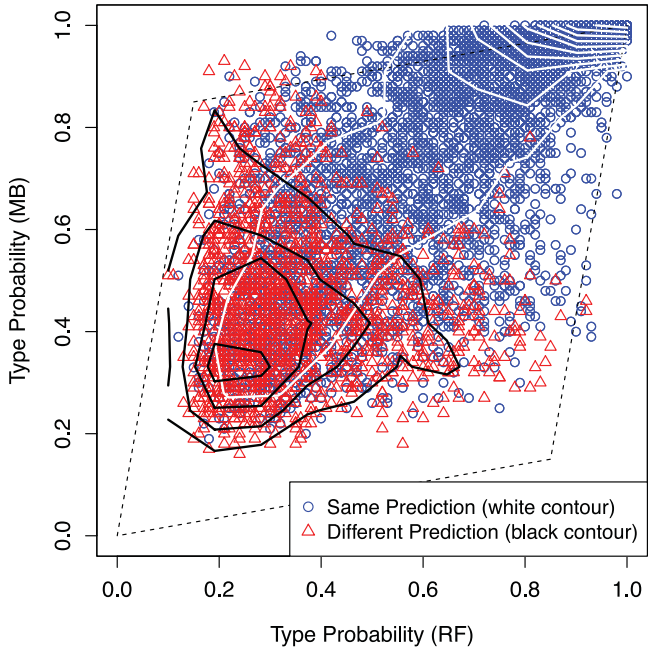

**Figure 8.** The probabilities $p$ of variability types predicted by random forests (RF) and multistage Bayesian networks (MB) are compared. About 74 per cent of the sources are associated with the same type by both classifiers (blue circles), while different types are predicted for the same source in about 26 per cent of the cases (red triangles). The reference dashed lines mark symmetric offsets with respect to the diagonal as $|p_{RF} - p_{MB}| = 0.7$ $(1 - |p_{RF} + p_{MB} - 1|)$.

**Table 6.** The number of predictions per variability type is presented for selected thresholds of the predicted type probability $p$, employing random forests and a multistage method based on Bayesian networks (BN). The total number of variability types predicted is 6051. Only types with at least one candidate with $p \geq 0.5$ are included in this table.

| Type | Random forests | | | Multistage BN | | |
|---|---|---|---|---|---|---|
| | $p$ 0.9 | $p$ 0.7 | $p$ 0.5 | $p$ 0.9 | $p$ 0.7 | $p$ 0.5 |
| I_X | – | – | 4 | 2 | 28 | 212 |
| LPV_P | – | – | 1 | – | – | 2 |
| LPV_X | 675 | 1900 | 2466 | 1520 | 2106 | 2389 |
| RS+BY_P | – | – | – | – | – | 2 |
| RS+BY_X | 19 | 158 | 451 | 25 | 177 | 455 |
| BE+GCAS_X | 8 | 71 | 209 | 70 | 182 | 301 |
| SPB_P | – | 4 | 14 | 5 | 49 | 112 |
| ACV_P | – | – | 9 | – | 12 | 44 |
| ACV_X | – | – | – | – | 9 | 48 |
| EA_P | – | 1 | 7 | – | 8 | 15 |
| EA_X | – | 33 | 97 | 30 | 75 | 135 |
| EB_P | – | – | 1 | – | 2 | 7 |
| ACYG_P | – | – | 1 | – | – | 3 |
| ACYG_X | – | 1 | 22 | – | 11 | 41 |
| BCEP_X | – | – | – | – | 14 | 51 |
| GDOR_P | – | – | – | – | – | 4 |
| GDOR_X | – | – | 9 | – | 9 | 35 |
| DSCT_P | – | – | – | – | – | 1 |
| DSCT_X | – | – | – | 3 | 62 | 124 |
| DSCTC_P | – | 4 | 20 | 2 | 20 | 42 |
| DSCTC_X | – | 43 | 150 | 7 | 127 | 228 |
| Total | 702 | 2215 | 3461 | 1664 | 2891 | 4251 |

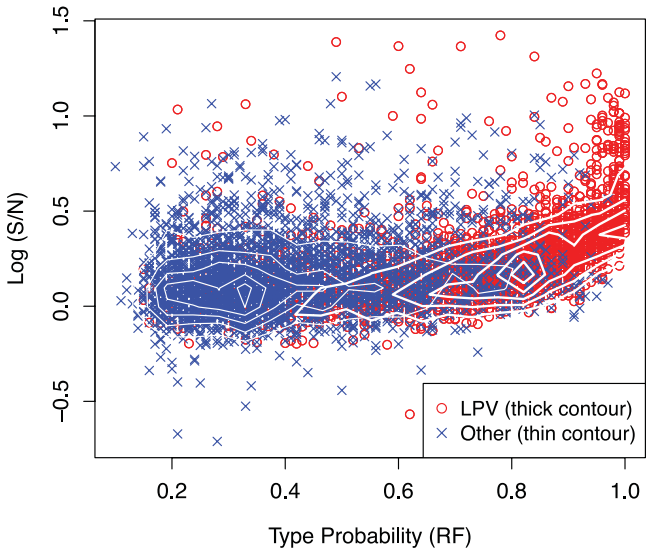

**Figure 9.** The S/N ratios of sources in the prediction set, plotted as a function of the probability of variability types (predicted by random forests), show that higher levels of S/N ratio and probability are populated mostly by the *LPV* types.

A further comparison of results with the literature is presented for those sources in the prediction set which are associated with a class, although uncertain, in the AAVSO catalogue (Watson et al. 2011). Two additional confusion matrices compare predicted versus uncertain AAVSO types[21] in Fig. 11: (a) the matrix on the left-hand side includes 1122 sources with any prediction probability, and (b) the matrix on the right-hand side is restricted to 998 objects with prediction probabilities greater than 0.5. For each source, the type predicted with the greatest probability, between random forests and the multistage classifiers, was selected. The comparison of such subsets of the two catalogues shows agreement at the level of 80 per cent in case (a) and of 87 per cent in case (b), after aggregating types related to eclipsing binaries (*E*, *EA*, *EB*) and Delta Scuti stars (DSCT, DSCTC).

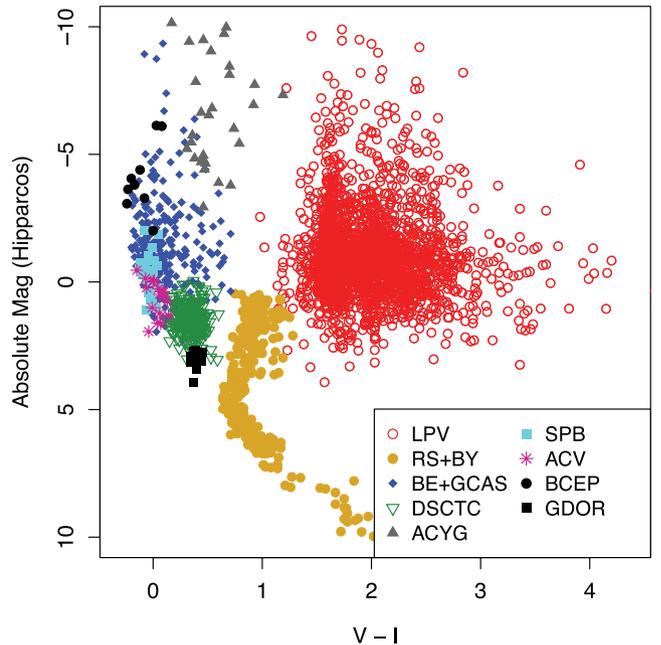

**Figure 10.** The Hertzsprung–Russell diagram includes stars with the same variability types predicted by both random forests and the multistage methodology involving Bayesian networks (with probabilities of 0.4 or greater according to both classifiers). Eclipsing binaries and irregular variables are excluded because they are not associated with specific colour–magnitude loci by definition. Limitations concerning the values of absolute magnitude (when parallax measurements are insignificant) are described in Section 5. The label suffixes 'P' and 'X' referring to the same types have been aggregated.

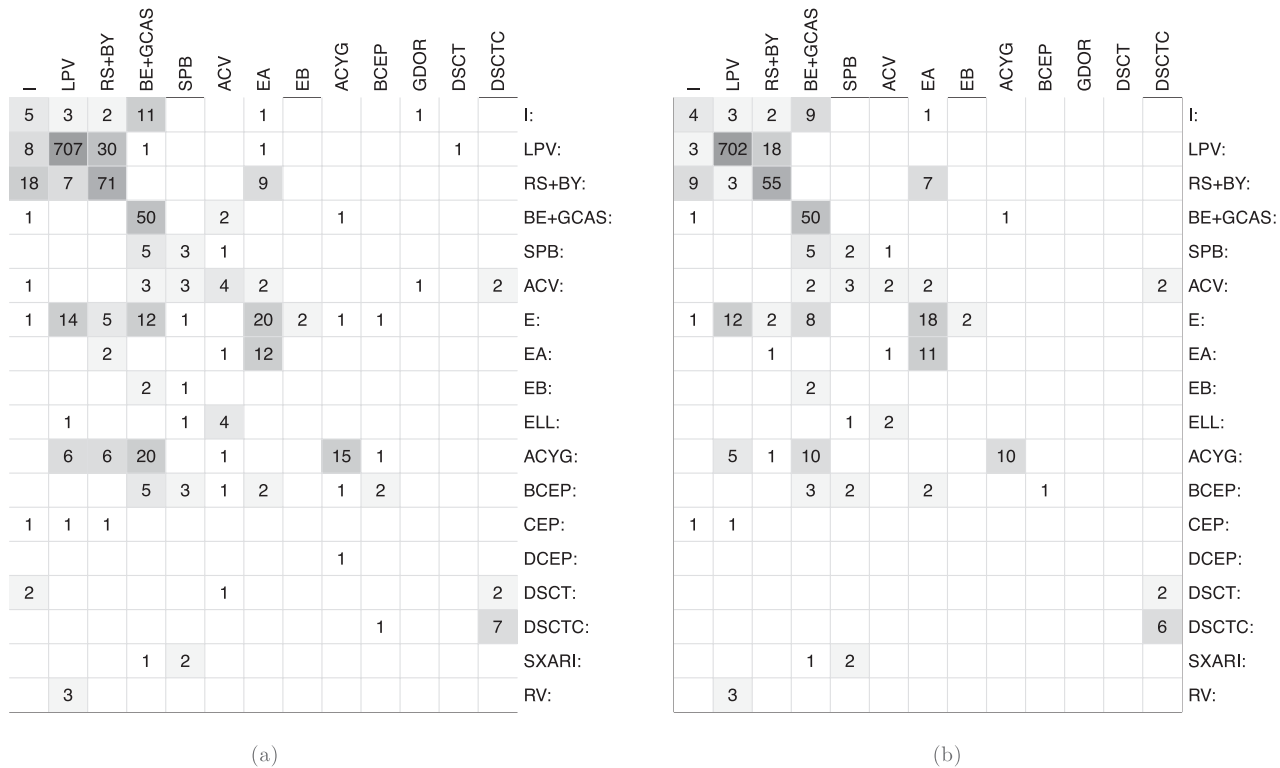

**Figure 11.** Comparison of a subset of predicted variability types (listed in columns) with uncertain classifications from the AAVSO catalogue (listed in rows) for any prediction probability in panel (a), and limited to prediction probabilities greater than 0.5 in panel (b).

A selection of periodic light curves of sources from the *Hipparcos* unsolved catalogue is shown in Fig. 12 with classification predictions and associated probabilities, illustrating a source with uncertain variability type available in the literature (Watson et al. 2011) and other objects which have been newly classified.

## 9 CONCLUSIONS

This paper extended the scope of automated classification of variable stars to objects manifesting irregular and non-periodic light variations. We employed the *Hipparcos* catalogue of unsolved variables to complement our training set with sources of irregular and non-periodic variability types, and predict types for thousands of stars flagged as variable but associated with uncertain or missing classification.

Our training set comprised 24 variability types and included 3881 sources with periodic, non-periodic and irregular light variations, which can be valuable to other multi-epoch surveys, with possible survey-specific adaptations.

Source properties and light-curve shapes were described by a set of attributes, which derived information from catalogues, modelling and statistical analyses. Attributes were selected and ranked owing to their importance for classification according to random forests. Less than 15 attributes were sufficient for characterizing both periodic and stochastic light variations with random forests.

An overall misclassification rate under 12 per cent was achieved, with confusion among variability types in agreement with expectations. Similar results were also obtained by a multistage methodology based on Bayesian networks.

The automated classification of the *Hipparcos* unsolved catalogue and a set of microvariables provided variability types for 6051 stars with uncertain or missing classification in the literature, employing both random forests and the multistage approach based on Bayesian networks. Despite the high levels of noise, sufficient information was still available to recognize several variability types, though caution was recommended in the interpretation of all classification results. Our predictions of variability types, complemented by the details of source characterization, are expected to be useful for selections of objects of interest and further investigations.

## ACKNOWLEDGMENTS

We would like to thank Joseph Richards for the detailed comments and constructive suggestions on the original manuscript. We are grateful to Elizabeth Waagen, Matthew Templeton and Sebastián Otero for clarifications regarding the AAVSO Variable Star Index. We thank Branimir Sesar for suggesting the consideration of the Galactic latitude as attribute. This paper makes use of the Hipparcos and Tycho Catalogues, which are the primary products of the European Space Agency's astrometric mission *Hipparcos*. LMS has been kindly supported by the Spanish Ministry of Economy under project number AyA2011-24052. TL has been kindly supported by the Austrian Science Fund under project number P20046-N16.

[21] The AAVSO classification uncertainty is indicated by a ':' symbol following type labels. The AAVSO notation has been simplified according to our class definitions in Table 1, and only the nearest classifications cross-matched within 1 arcsec from the *Hipparcos* sources have been considered. Alternative or generic (e.g. VAR, PER, PULS, MISC) variability types from the AAVSO catalogue and a few rare types not present in our training set have not been included in Fig. 11. Additional GCVS type labels from the literature are listed in Table 7 of the online supporting information, together with other commonly used symbols.

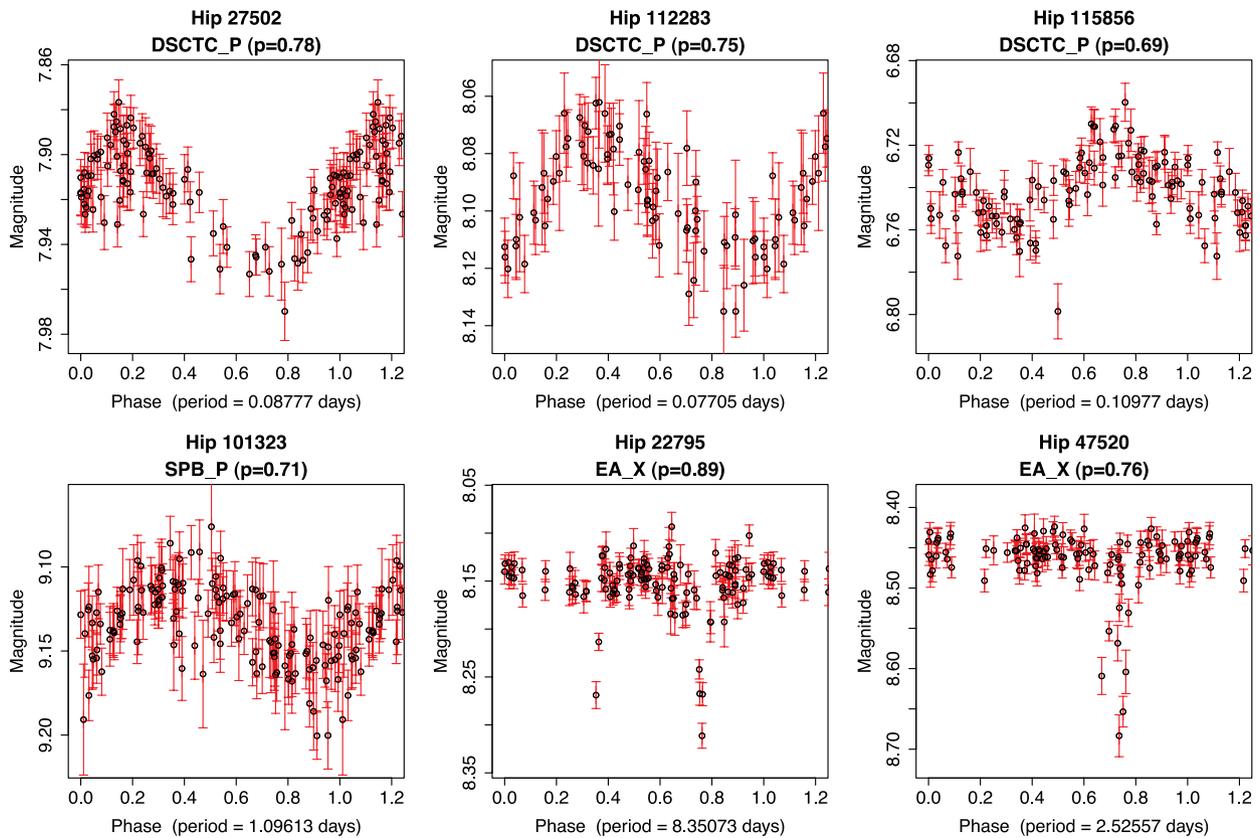

**Figure 12.** A selection of periodic light curves of sources from the *Hipparcos* unsolved catalogue is shown with classification predictions and associated probabilities (from random forests), as indicated in each panel. Among the sources illustrated in this figure, only Hip 22795 has a variability type available in the literature (uncertain *E:*, according to Watson et al. 2011), while the other objects are newly classified. The periods corresponding to eclipsing binaries (Hip 22795 and Hip 47520) have been doubled, as explained in Section 5.2.

## APPENDIX A: ATTRIBUTE DISTRIBUTIONS

The distributions of values of the 15 most important attributes in the training set (described in Section 5) are shown in Figs A1–A3 for each variability type.

## APPENDIX B: THE MULTISTAGE TREE

The multistage classification tree derived as explained in Section 7 is shown in Fig. B1.

## APPENDIX C: PREDICTION PROBABILITIES

The classifier models returned a probability array for each source of the prediction set, measuring the likelihood of association with all of the variability types included in the training set. The full classification probability arrays are available only online from the Centre de Données astronomiques de Strasbourg website (http://cdsarc.u-strasbg.fr). The online Tables C1 and C2 (see also the supporting information) refer to the probabilities obtained from random forests and from the multistage methodology involving Bayesian networks, respectively.

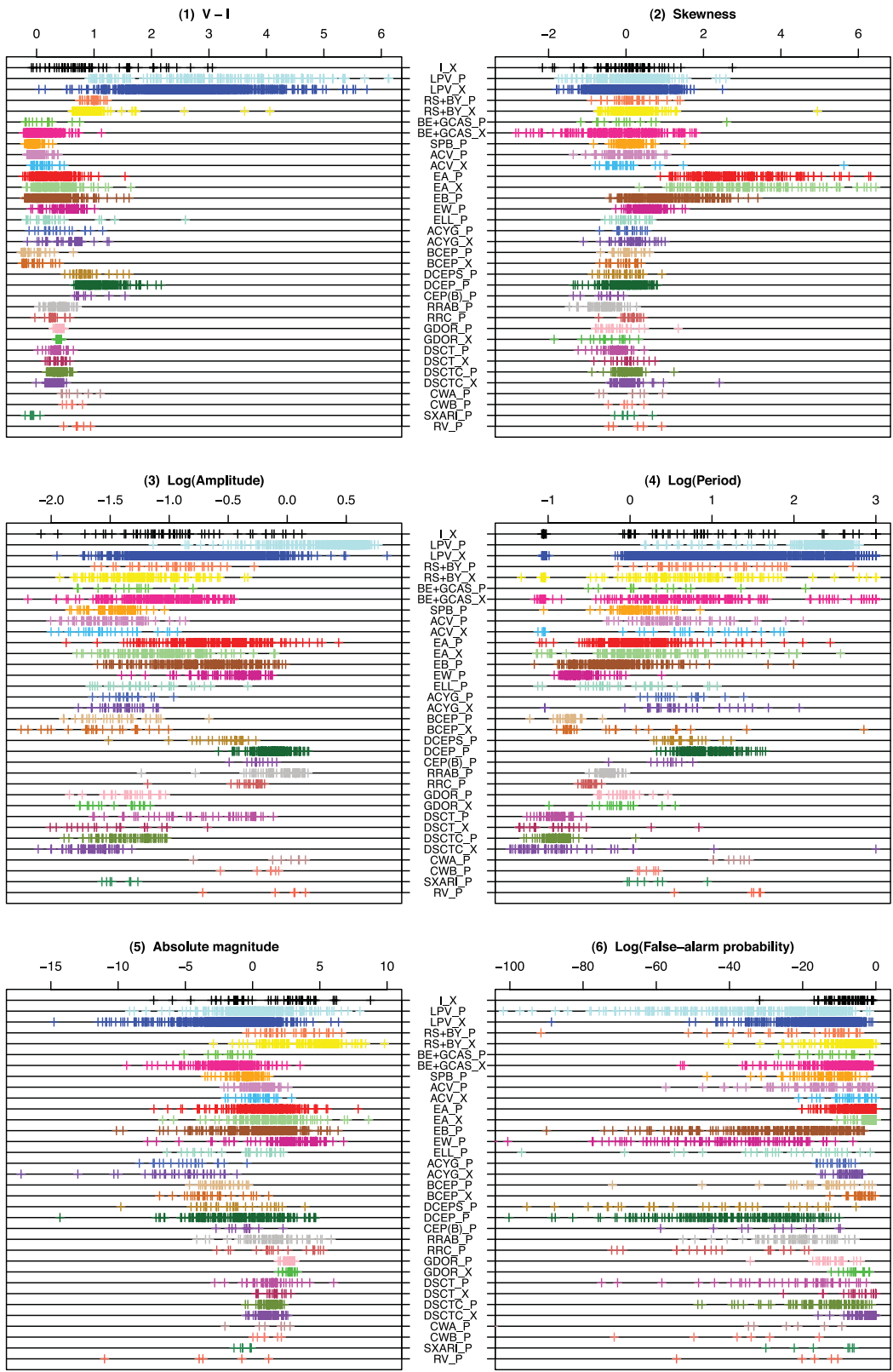

**Figure A1.** The distribution of values of the first six most important attributes in the training set for each variability type.

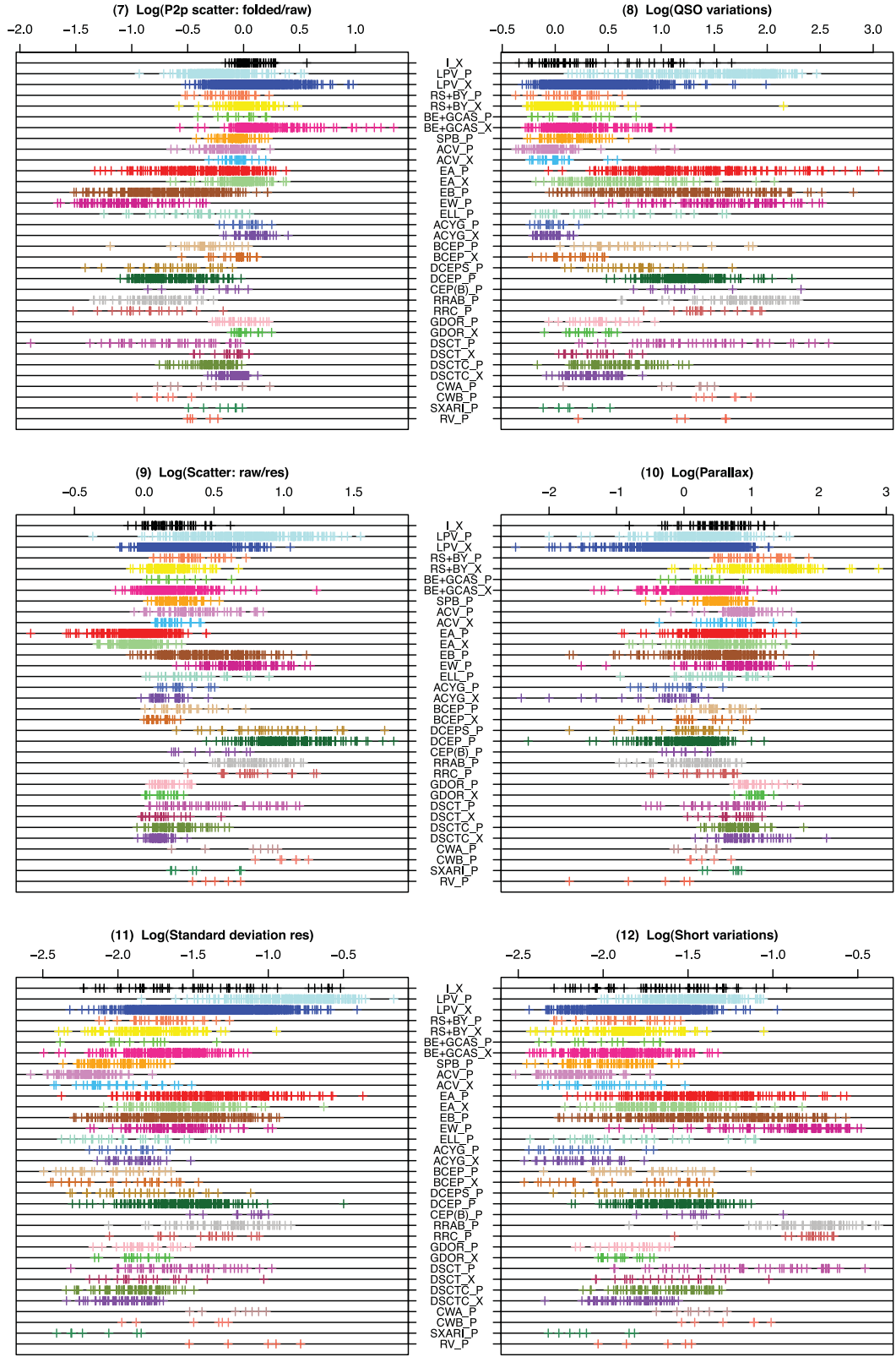

**Figure A2.** The distribution of values of the 7th to the 12th most important attributes in the training set for each variability type.

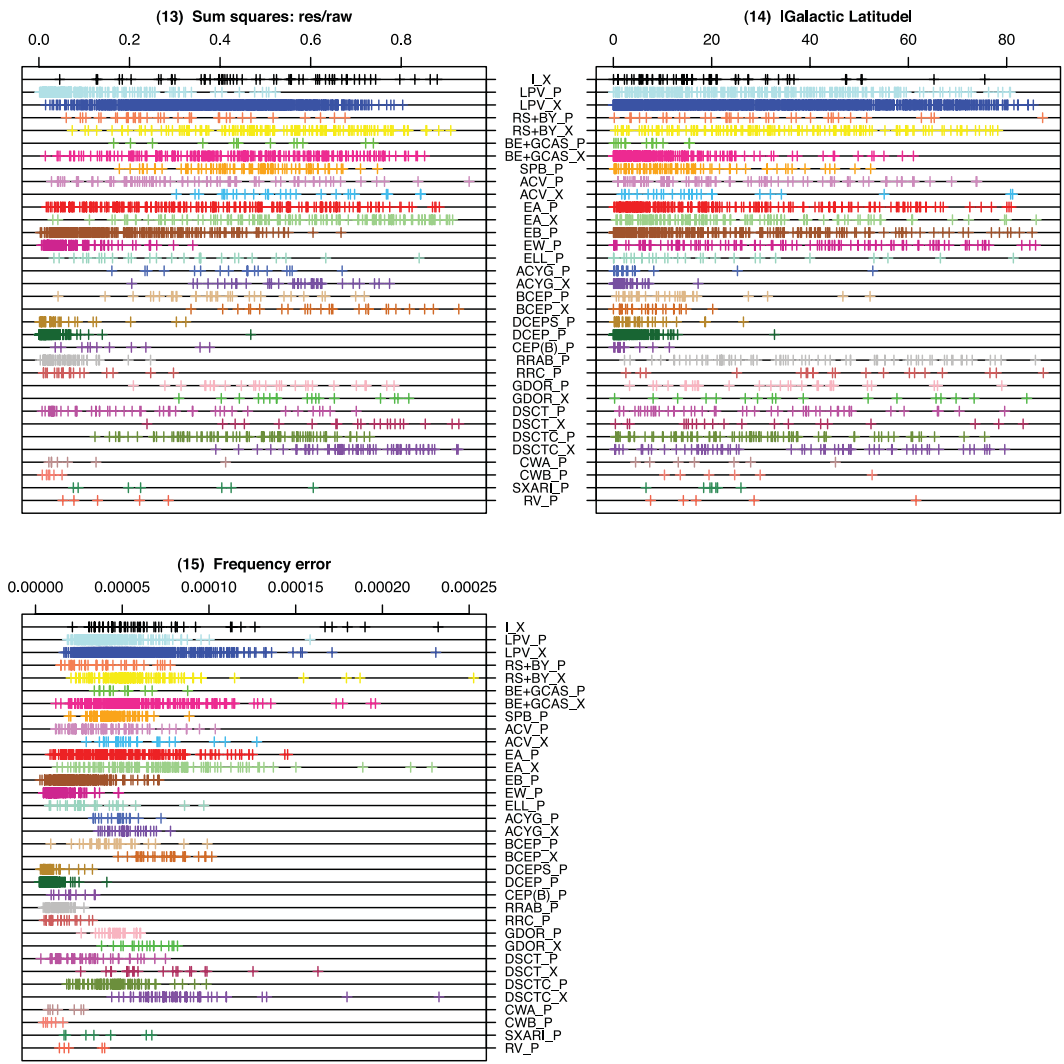

**Figure A3.** The distribution of values of the 13th to the 15th most important attributes in the training set for each variability type.

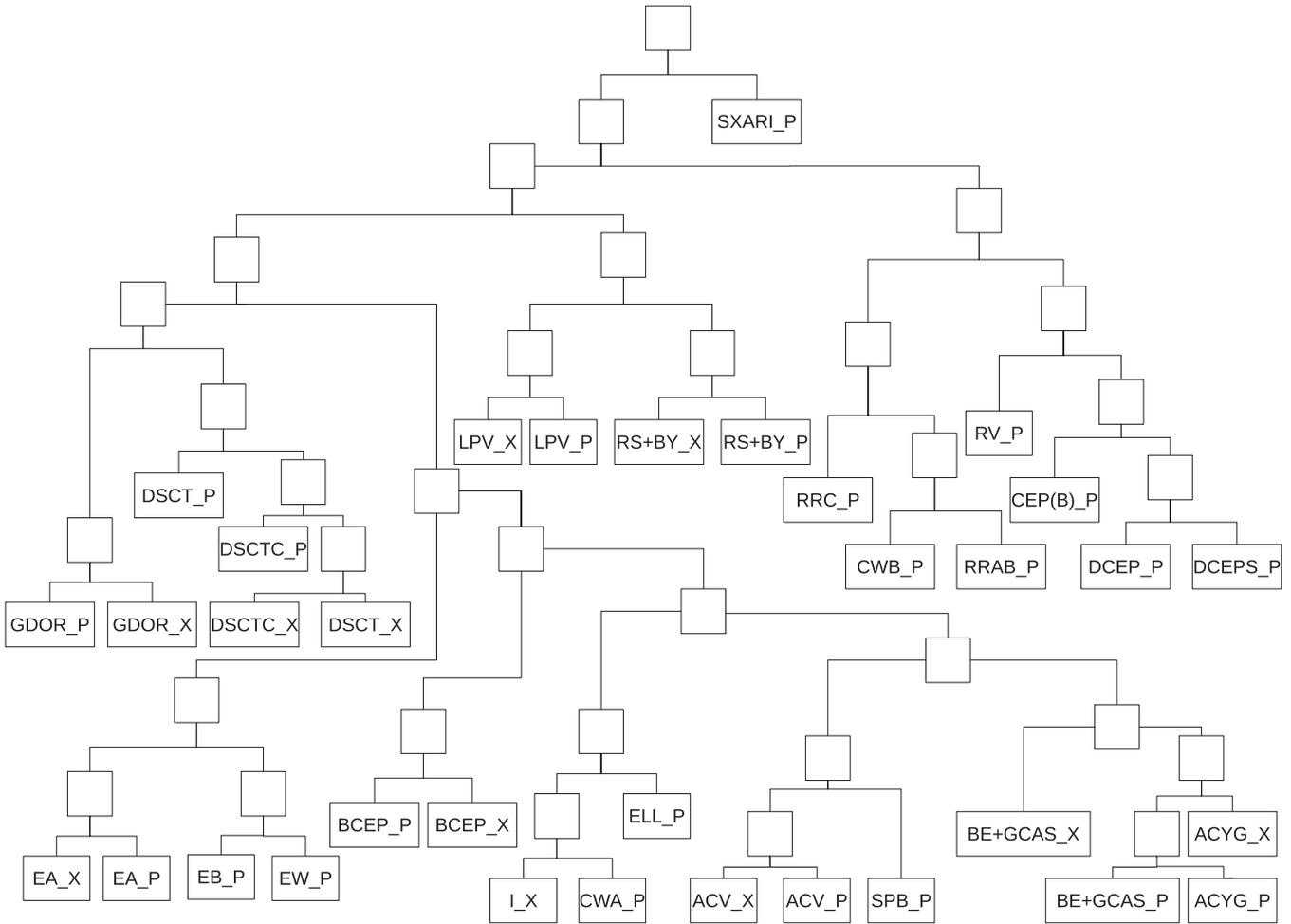

**Figure B1.** Multistage tree for the classifier described in Section 7. The empty squares represent classification nodes (and the group of types that they handle).

## SUPPORTING INFORMATION

Additional Supporting Information may be found in the online version of this paper:

**Table 2.** The data refer to the training-set stars selected from the *Hipparcos* periodic, unsolved and microvariable sets.
**Table 4.** The data refer to the set of stars for which variability types are predicted, selected from the *Hipparcos* unsolved catalogue and a set of microvariables.
**Table 5.** Variability types predicted by random forests (RF) and from a multistage method based on Bayesian networks (MB) are listed together with the *Hipparcos* identifiers, the *Hipparcos* sets from which the sources have been selected ($U_1$, $U_2$ and M, as defined in Section 3), and literature variability types from the *Hipparcos* and the AAVSO catalogues when available.

**Table 7.** List of additional symbols and labels for variability types from the literature, which are included together with our classification results where available.
**Tables C1 and C2.** The full probability arrays of predictions obtained from random forests and from the multistage methodology involving Bayesian networks, respectively.

Please note: Wiley-Blackwell are not responsible for the content or functionality of any supporting materials supplied by the authors. Any queries (other than missing material) should be directed to the corresponding author for the paper.

This paper has been typeset from a TeX/LaTeX file prepared by the author.